\documentclass[12pt]{article}
\DeclareMathAlphabet{\scr}{U}{rsfs}{m}{n}
\usepackage{latexsym}
\usepackage[mathscr]{eucal}
\usepackage{amsfonts}
\usepackage{amscd}
\usepackage{cite}
\usepackage{graphicx}
\usepackage{amssymb}
\usepackage[centertags]{amsmath}
\usepackage{dsfont}
\usepackage{yfonts}
\usepackage{enumerate}

\newcommand{\cleqn}{\setcounter{equation}{0}}
\newcommand{\newc}{\newcommand}
\newc{\be}{\begin{equation}}
\newc{\ee}{\end{equation}}
\newc{\bea}{\begin{eqnarray}}
\newc{\eea}{\end{eqnarray}}
\newc{\ol}{\overline}
\newc{\wt}{\widetilde}
\newc{\bs}{\boldsymbol}
\newc{\m}{\mathcal}
\newc{\vl}{\langle}
\newc{\vr}{\rangle}
\begin{document}

\title{\hfill ~\\[-30mm]
       \hfill\mbox{\small SHEP-09-30}\\[30mm]
        \textbf{A Supersymmetric Grand Unified Theory of Flavour with $\bs{PSL_2(7)\times SO(10)}$}}
\date{}
\author{\\Stephen F. King\footnote{E-mail: {\tt king@soton.ac.uk}}~~and
        Christoph Luhn\footnote{E-mail: {\tt christoph.luhn@soton.ac.uk}}\\ \\
  \emph{\small{}School of Physics and Astronomy, University of Southampton,}\\
  \emph{\small Southampton, SO17 1BJ, United Kingdom}}

\maketitle

\begin{abstract}
\noindent
We construct a realistic Supersymmetric Grand Unified Theory of Flavour based on $PSL_2(7)\times SO(10)$,
where the quarks and leptons in the {\bf 16} of $SO(10)$
are assigned to the complex triplet representation
of $PSL_2(7)$, while the flavons are assigned to a combination of sextets
and anti-triplets of $PSL_2(7)$. Using a $D$-term vacuum alignment mechanism,
we require the flavon sextets of $PSL_2(7)$ to be aligned along the 3-3
direction leading to the third family Yukawa couplings, while the flavon
anti-triplets describe the remaining Yukawa couplings. Other sextets are
aligned along the neutrino flavour symmetry preserving directions leading to
tri-bimaximal neutrino mixing via a type II see-saw mechanism, with
predictions for neutrinoless double beta decay and cosmology. 
\end{abstract}

\vspace{30mm}

~~~Keywords: discrete family symmetries, tri-bimaximal mixing, GUT

\thispagestyle{empty}
\vfill
\newpage
\setcounter{page}{1}

\section{Introduction}
\cleqn

The discovery of neutrino mass and approximately tri-bimaximal (TB) lepton
mixing \cite{HPS} suggests some kind of a non-Abelian discrete family symmetry
$G_f$ might be at work, at least in the lepton sector. In the neutrino flavour
basis (i.e. diagonal charged lepton mass basis), it has been shown that the TB
neutrino mass matrix is invariant under $S,U$ transformations,
${M^{\nu}_{TB}}\,= S {M^{\nu}_{TB}} S^T\,= U {M^{\nu}_{TB}} U^T $
\cite{King:2009mk}. A very straightforward argument shows that this neutrino
flavour symmetry group has
only four elements corresponding to Klein's four-group $Z_2^S \times
Z_2^U$. By contrast the diagonal charged lepton mass matrix (in this basis)
satisfies a diagonal phase symmetry $T$. The matrices $S,T,U$ form the
generators of the group $S_4$ in the triplet representation, while the $A_4$
subgroup is generated by $S,T$. The observed neutrino flavour symmetry
corresponding to the two generators $S,U$ may arise either directly or
indirectly from a range of discrete symmetry groups
\cite{King:2009ap}. Examples of the direct approach, in which one or more
generators of the discrete family symmetry appears in the neutrino flavour
group, are typically based on $S_4$ \cite{Lam:2009hn} or a related group such
as $A_{4}$ \cite{Ma:2007wu,Altarelli:2006kg} or $PSL_2(7)$
\cite{King:2009mk}. Models of the indirect kind, in which the neutrino flavour
symmetry arises accidentally, include also $A_4$ \cite{King:2006np} and $S_4$
\cite{Dutta:2009bj} as well as $\Delta_{27}$ \cite{deMedeirosVarzielas:2006fc}
and the continuous flavour symmetries like, e.g., $SO(3)$ \cite{King:2006me}
or $SU(3)$ \cite{King:2003rf} which accommodate the discrete groups above as
subgroups \cite{deMedeirosVarzielas:2005qg}. For an incomplete list of models
with family symmetries see
\cite{S3-L,Dn-L,A4-L,S4-L,delta54-L,A5-L,S3-LQ,Dn-LQ,Q6-LQ,A4-LQ,A4-SU5,doubleA4-LQ,S4-LQ,Z7Z3-LQ,T7-LQ} and the reviews \cite{Reviews,Review-inducedVEV}.

A desirable feature of a complete model of quark and lepton masses and mixing
angles is that it should be consistent with an underlying Grand Unified Theory
(GUT) structure, either at the field theory level or at the level of the
superstring. The most ambitious models which have been built to achieve this
are based on an underlying $SO(10)$ structure. This is very constraining
because it requires that all the 16 spinor components of a single family
should have the same family charge, comprising the left-handed fermions $\psi$
and the CP conjugates of the right-handed fermions $\psi^c$, including the
right-handed neutrino. Although it seems somewhat {\it ad hoc} to add
right-handed neutrino singlets to $SU(5)$, once they are present it is
straightforward to construct models of lepton masses and TB mixing that are
consistent with quark masses and mixing, and there are several successful
models of this kind for example based on $A_4\times SU(5)$
\cite{A4-SU5}.

Despite the theoretical attractiveness of $SO(10)$, there are very few
$SO(10)$ models capable of enforcing TB mixing by means of a family
symmetry. It is desirable that such models contain complex triplet
representations of the family symmetry and  examples of such models based on
the Pati-Salam subgroup of $SO(10)$ have been constructed where the family
group is $SU(3)_{f}$ \cite{King:2003rf} or $\Delta_{27}$
\cite{deMedeirosVarzielas:2006fc}, with Yukawa couplings arising from
operators of the
form $\bar{\phi}^{i}\psi^{}_{i}\bar{\phi}^{j}\psi_{j}^{c}$. If the triplet
representations were taken to be real rather than complex then this would
allow an undesirable alternative contraction of the indices in the Yukawa
operator namely $\psi_{i}\psi_{i}^{c}\bar{\phi}^{j}\bar{\phi}^{j}H$ leading to
a Yukawa matrix proportional to the unit matrix which would tend to destroy
any hierarchies in the Yukawa matrix. For real representations, there is no
symmetry at the effective operator level that could forbid such a trivial
contraction of the two triplet fermion fields. However in principle it is
possible to appeal to the details of the underlying theory in order to forbid
the trivial contraction. This involves a discussion of the heavy messenger
states whose exchange generates the operator, for example as was done recently
in the $S_4 \times SO(10)$ model in \cite{Dutta:2009bj}.

A general problem with all $SO(10)$ models in which the quarks and leptons
form triplet representations of the family symmetry (necessary to account for
TB mixing) is that the top quark Yukawa coupling arises from a double flavon
suppressed operator of the form
$\bar{\phi}^{i}\psi^{}_{i}\bar{\phi}^{j}\psi_{j}^{c}$. The situation improves
somewhat if one considers flavons also in sextet representations, as pointed
out in \cite{King:2009mk}. For example, introducing a two index anti-sextet of
$SU(3)_f$, $\hat{\chi}^{ij}$, the lowest order Yukawa operators become,
$\hat{\chi}^{ij}\psi^{}_{i}\psi_{j}^{c}$. If the anti-sextet flavon
$\hat{\chi}$ has a VEV aligned along the 3-3 direction $\langle
\hat{\chi}^{ij} \rangle = V\delta_{i3}\delta_{j3}$, then this operator with a
coefficient $y/M$ would imply a third family Yukawa coupling of $yV/M$ with
the top quark Yukawa coupling of 0.5 implying $V /M \approx 0.5y^{-1}$, which
has acceptable convergence properties.

Flavon sextets are therefore well motivated as the origin of the third
family Yukawa couplings in $SO(10)$ models. Although this is possible in the
case where the family group is $SU(3)_{f}$ \cite{King:2003rf}, it is not
possible for the discrete groups $\Delta_{27}$
\cite{deMedeirosVarzielas:2006fc}
or $S_4$ \cite{Dutta:2009bj} for the simple reason that these groups
do not admit sextet representations.  The smallest simple discrete group which
contains complex triplets and sextet representations is ${PSL_2(7)}$, which is
the projective special linear group of two dimensional matrices over the
finite Galois field of seven elements. ${PSL_2(7)}$ contains 168 elements and
is sometimes written as $\Sigma(168)$ \cite{FFK}. The relationship of
${PSL_2(7)}$ to some other family symmetries that have been used in the
literature is discussed in \cite{Z7Z3-LQ,PSLgroup,discAnom}.

In a recent paper \cite{King:2009mk} we developed the representation theory of
${PSL_2(7)}$ for triplets and sextets in a convenient basis suitable for
applications of ${PSL_2(7)}$ as a family symmetry capable of describing quark
and lepton masses and mixing angles in the framework of $SO(10)$ models. We
showed how the triplet representation given in terms of the standard
generators $A,B$ in \cite{PSLgroup} may be related to four ${PSL_2(7)}$
generators $S,T,U,V$. In such a basis the subgroup structure ${PSL_2(7)}
\supset S_4 \supset A_4$ just corresponds to the respective generators being
$S,T,U,V \supset S,T,U \supset S,T$.

The purpose of the present paper is to construct a realistic Supersymmetric
Grand Unified Theory of Flavour based on $PSL_2(7)\times SO(10)$. We require 
the sextets to be aligned along the 3-3 direction to account for the large
third family Yukawa couplings, while we shall make use of anti-triplet
flavons, whose VEVs are aligned along the columns of the TB mixing matrix, to
account for the first and second family quark and lepton masses and mixings. 
It turns out that in ${PSL_2(7)}$ it is easier to obtain the sextet flavon
vacuum alignments first using the $D$-term approach to vacuum alignment
discussed in \cite{King:2009ap}, but here applied to flavon sextets rather
than flavon triplets. 
In this way we can obtain sextet vacuum alignments along the
3-3 direction, suitable for the top quark Yukawa coupling, if we assume two
relations (possibly arising by virtue of a higher symmetry) amongst different
quartic sextet combinations appearing in the flavon potential.
Sextet vacuum alignments along the $S,U$ preserving directions, suitable for
reproducing the neutrino flavour symmetry of TB mixing in a direct way as a
subgroup of ${PSL_2(7)}$, can be naturally obtained.
Once the sextet flavons have been aligned, the anti-triplet
flavons are then aligned against the pre-aligned sextet flavons, in particular
using the $S,U$ preserving sextet flavons, leading to triplet flavon
alignments along the columns of the TB mixing matrix as mentioned.

\begin{figure}
\begin{center}
\includegraphics[width=7cm]{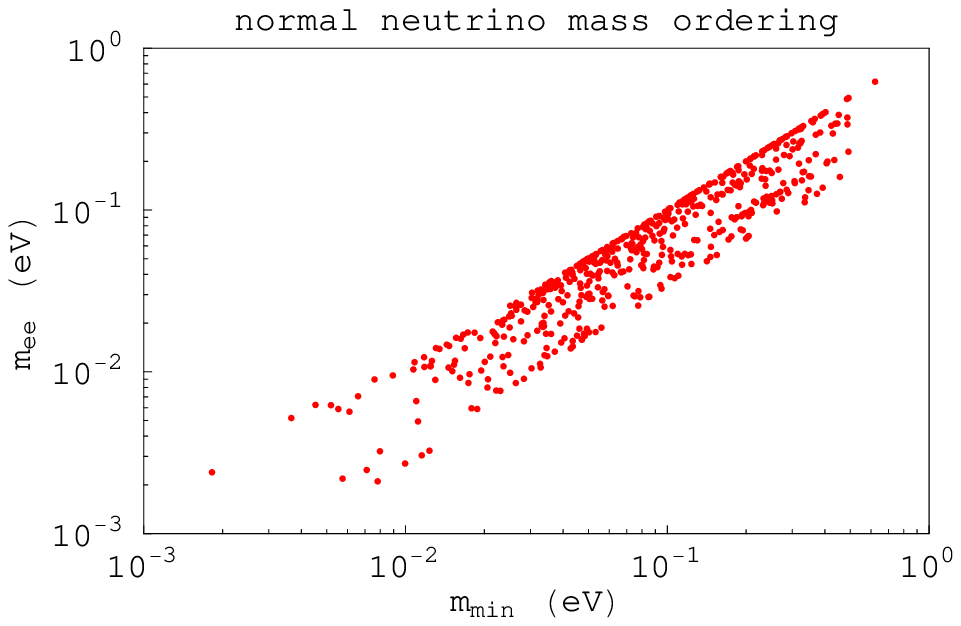} \qquad
\includegraphics[width=7cm]{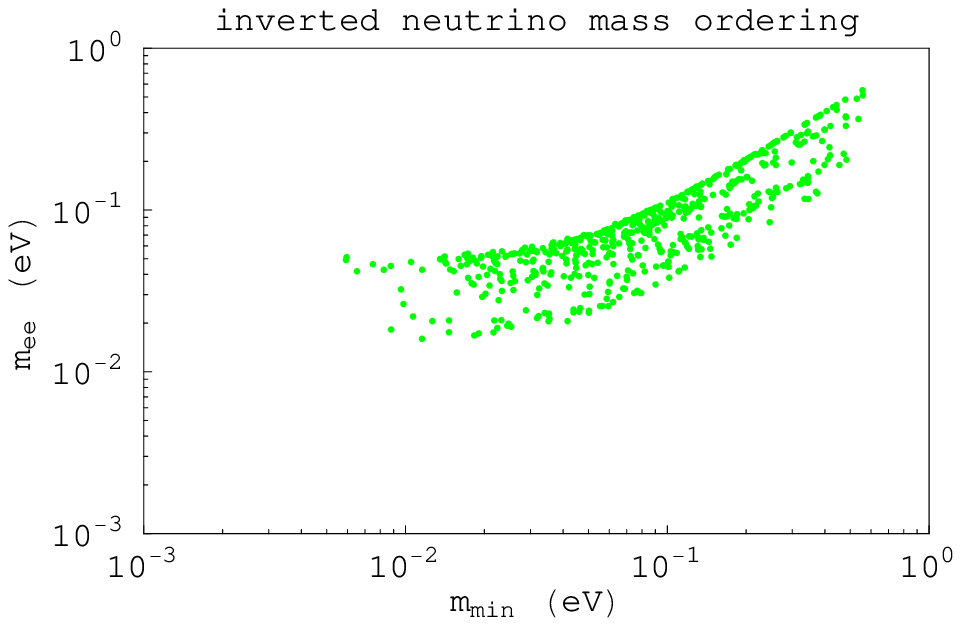}  \\[5mm]
\includegraphics[width=7cm]{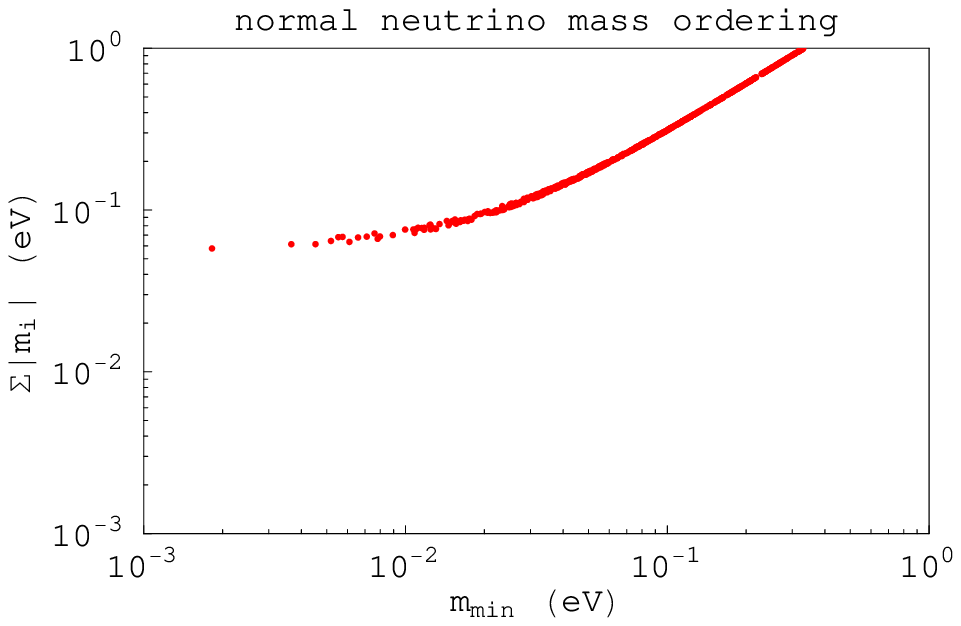} \qquad
\includegraphics[width=7cm]{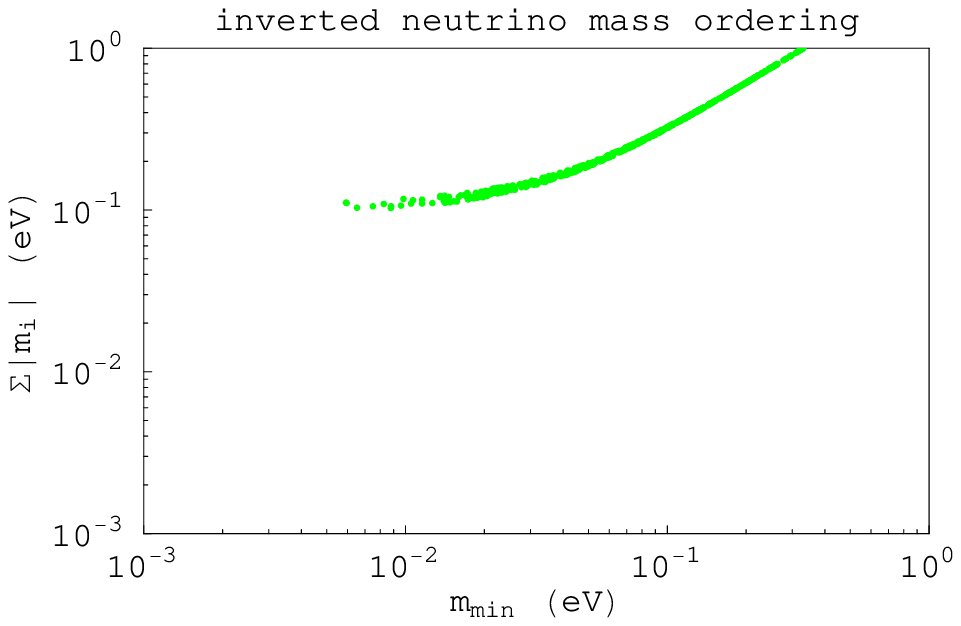}
\end{center}
\caption{\label{fig-Mee-Mmin-scatterLOG}\small Scatter plots for $m_{ee}$ and $\sum_i
  |m_i|$  against $m_{\mathrm{min}}$ for the type II see-saw model
  based on $PSL_2(7)\times SO(10)$.}
\end{figure}

In general the neutrino mass and mixing in this model can arise from either
the type I see-saw or the type II see-saw or both. The discussion of the type
I see-saw approach follows along the lines of the models in
\cite{King:2003rf,deMedeirosVarzielas:2006fc} based on constrained
sequential dominance
\cite{King:1998jw}, leading to the indirect type of models, since the triplet
flavon aligned along the third column of the TB mixing matrix breaks both $S$
and $U$. Here we shall focus on
the new possibility offered by the $PSL_2(7)\times SO(10)$ model of the type
II see-saw mechanism where the sextet flavons aligned along the $S,U$
preserving directions enter the neutrino sector, thereby preserving these
generators leading to the neutrino flavour symmetry being reproduced in a
direct way. As a preview of our results, we shall find that the type II
see-saw model leads to TB mixing in the neutrino sector with a mass spectrum
spanning the range from hierarchical or inverse hierarchical, up to the
quasi-degenerate region. The resulting predictions for neutrinoless double
beta decay mass parameter $m_{ee}$ and the total sum of physical neutrino
masses  $\sum_i|m_i|$ relevant for cosmological hot dark matter are both shown
in Fig.~\ref{fig-Mee-Mmin-scatterLOG} as a function of the lightest physical
neutrino mass $m_{\mathrm{min}}$ using a double logarithmic scale. These results are in
sharp contrast to the case of the type I see-saw possibility with constrained
sequential dominance which would lead to a strong neutrino mass hierarchy with
$m_{\mathrm{min}}= |m_1| \sim 0$. Later on we shall analyse the type II see-saw results
in more detail using linear scales.

The layout of the remainder of the paper is as follows.
In Section~2 we present the symmetries and superfield content of the
SUSY $PSL_2(7)\times SO(10)$ model and
discuss the desired vacuum alignments and the resulting fermion mass matrices.
In Section~3 we discuss the type II neutrino phenomenology in the model.
Section~4 is devoted to a detailed analysis of the $D$-term vacuum alignment for the
sextet and ant-triplet flavons. Section~5 concludes the paper.

\section{Fermion masses in the $\bs{PSL_2(7)\times SO(10)}$ model}
\cleqn
\subsection{The desired vacuum alignments}
As discussed in the Introduction, the sextet flavons are used in two ways in
the model. The sextet $\hat{\chi}^{ij}_{\mathrm{top}}$ aligned along the 3-3
direction is responsible for the third family Yukawa couplings, including that
of the top quark, via operators like (dropping the Higgs fields)
\be
\psi^{}_{i}\hat{\chi}^{ij}_{\mathrm{top}}\psi_{j}^{c} \ .
\label{opYuk-2}
\ee
The sextets $\hat{\chi}^{ij}_{TB}$ aligned along the $S$ and $U$ preserving
directions are responsible for the effective Majorana couplings, via operators
like (again dropping the Higgs fields)
\be
\psi^{}_{i}\hat{\chi}^{ij}_{TB}\psi^{}_{j} \  + \psi^{c}_{i}\hat{\chi}^{ij}_{TB}\psi^{c}_{j} ,
\label{opMaj-2}
\ee
where the first term above contributes to the type II see-saw mass while the
second term is responsible for the heavy right-handed neutrino masses. As
discussed in \cite{King:2009mk}, the sextet flavon 3 by 3 matrices
$\hat{\chi}_{}^{}$ which enter the above couplings to quarks and leptons, are
related to the six component column vector
\be
\chi = (\chi_1, \chi_2,\chi_3,\chi_4,\chi_5,\chi_6)^T
\ee
by,
\bea
\hat \chi&\!\!=\!\!& - \frac{(1+i)}{6\sqrt{2}}
\left[
\chi_1 \begin{pmatrix}
4&1&1\\1&-2&-2\\1&-2&-2
\end{pmatrix}
-i\sqrt{3}\,\chi_2 \begin{pmatrix}
0&1&-1\\1&2&0\\-1&0&-2
\end{pmatrix}
-i\sqrt{3}\,b_7 \,\chi_3 \begin{pmatrix}
0&1&-1\\1&-1&0\\-1&0&1
\end{pmatrix}
\right.\nonumber\\[2mm]
&&+ \left.\sqrt{3}\,b_7 \,\chi_4 \begin{pmatrix}
0&1&1\\1&1&0\\1&0&1
\end{pmatrix}
+ \sqrt{2} \,\chi_5 \begin{pmatrix}
2&-1&-1\\-1&2&-1\\-1&-1&2
\end{pmatrix}
- i \sqrt{6}\,\bar b_7 \,\chi_6 \begin{pmatrix}
1&0&0\\0&0&1\\0&1&0
\end{pmatrix}
\right] ,\label{yukmatrixreal}
\eea
where we adopt the notation of the ``Atlas of finite groups'' \cite{atlas}
which defines
\be
b_7~=~\frac{1}{2} (-1+i\sqrt{7}) \ , \qquad
\bar b_7~=~\frac{1}{2} (-1-i\sqrt{7}) \ .
\ee
Mass matrices which are of the tri-bimaximal form are obtained from
alignments of the form
\be
\langle \chi^{}_{{TB}} \rangle ~=~
(0\,,\,0\,,\,0\,,\,  \alpha_4 \,,\,
\alpha_5 \,,\,\alpha_6)^T \ . \label{TBalignment}
\ee
The fully realistic type II model will require three flavons of this type
which we label as $\chi^{[p]}_{TB}$, where $p=0,1,2$. Explicitly their
alignments read
\bea
\langle \chi^{[0]}_{{TB}} \rangle &\propto&  (0\,,\,0\,,\,0\,,\,  0 \,,\,
0 \,,\,1 )^T \ , \label{tb-0-al}\\
\langle \chi^{[1]}_{{TB}} \rangle &\propto& \frac{1}{6} \cdot (0\,,\,0\,,\,0\,,\,  -\sqrt{14} \,,\,
-\sqrt{21} \,,\,-1 )^T \ , \label{tb-1-al}\\
\langle \chi^{[2]}_{{TB}} \rangle &\propto& \frac{1}{6} \cdot (0\,,\,0\,,\,0\,,\,  -\sqrt{14} \,,\,
\sqrt{21} \,,\,-1 )^T \ .\label{tb-2-al}
\eea
In order to generate a Yukawa matrix which gives mass to only the third
generation (top quark, bottom quark or tau lepton), we need the alignment
\be
\langle \chi^{}_{\mathrm{top}} \rangle ~\propto ~
\frac{(1-i)}{3\sqrt{2}} \cdot  (1~,~i \sqrt{3}~,~ -i\sqrt{3}/b_7 ~,~ -\sqrt{3}/b_7~,~-\sqrt{2} ~,~0)^T \ .\label{topalignment}
\ee
To obtain the alignments of Eqs.~(\ref{tb-0-al}-\ref{topalignment}) it is
necessary to study the $PSL_2(7)$ symmetric potential for the flavon sextet,
which we postpone to Section \ref{vac-sec}.

In addition the model also relies on the anti-triplet flavon fields
$\bar\phi_{23}$ and $\bar\phi_{123}$ whose VEVs become aligned along the
directions
\be
\langle \bar\phi_{23} \rangle
~\propto~\frac{1}{\sqrt{2}}\cdot\begin{pmatrix}0\\1\\-1\end{pmatrix} \ ,
\qquad
\langle \bar\phi_{123} \rangle
~\propto~\frac{1}{\sqrt{3}}\cdot\begin{pmatrix}1\\1\\1\end{pmatrix} \ ,
\label{anti-tri-ali}
\ee
where again we postpone discussion of these alignments until later.

\subsection{The symmetries and operators of the model}
\begin{figure}
\begin{center}
\includegraphics[width=15cm]{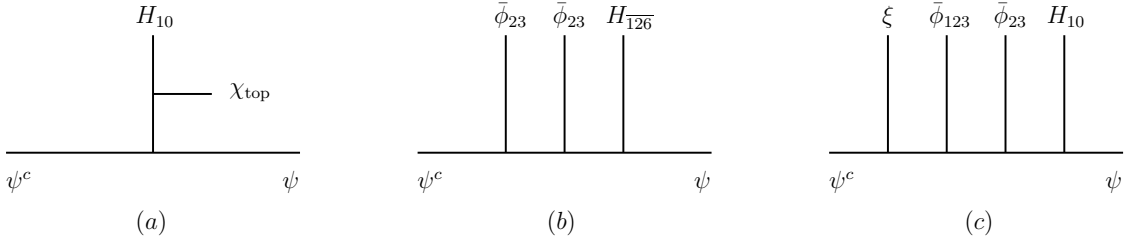}
\caption{\label{yuk-fig}\small The diagrams which generate the Yukawa
  couplings. In Pati-Salam language, $SU(4)\times SU(2)_L \times SU(2)_R$, the
  relevant components of the $SO(10)$ representations are:
  $\psi \rightarrow ({\bf 4},{\bf 2},{\bf 1})$,
  $\psi^c \rightarrow ({\bf \ol 4},{\bf 1},{\bf 2})$,
  $H_{10 } \rightarrow ({\bf 1},{\bf 2},{\bf 2})$,
  $H_{\ol{126}} \rightarrow ({\bf 15},{\bf 2},{\bf 2})$.}
\end{center}
\end{figure}

Having indicated the desired vacuum alignment of the flavon fields
$\chi^{}_{\mathrm{top}}$, $\chi^{[p]}_{TB}$, $\bar\phi^{}_{23}$,
$\bar\phi^{}_{123}$,
we can now formulate an $SO(10)$ model of fermion
masses using the $PSL_2(7)$ family symmetry. Table~\ref{particle-content}
lists the particle content together with all transformation properties.
\begin{table}
\begin{center}
$
\begin{array}{|c|c|c|c|c|c|c|c|c|c|}\hline
{}_{\phantom{{\big{|}}}}
\text{field}{}^{\phantom{{\big{|}}}} &
\psi & H_{10} & H_{\ol{126}}  & \Delta_{\ol{126}} &
\chi^{}_{\mathrm{top}} & \chi^{[p]}_{TB} &
\bar \phi_{23} & \bar \phi_{123}&\xi \\ \hline
{}_{\phantom{{\big{|}}}}SO(10){}^{\phantom{{\big{|}}}} &
{\bf 16} &  {\bf 10} & {\bf \ol{126}} &  {\bf \ol{126}} &
{\bf 1} &{\bf 1} &
{\bf 1} &{\bf 1} &{\bf 1} \\ \hline
{}_{\phantom{{\big{|}}}}PSL_2(7){}^{\phantom{{\big{|}}}} &
{\bf 3} & {\bf 1} &{\bf 1} &{\bf 1} &
{\bf 6} & {\bf 6} & {\bf \ol{3}} & {\bf \ol{3}} & {\bf 1} \\\hline
{}_{\phantom{{\big{|}}}}U(1){}^{\phantom{{\big{|}}}} &
0&1&4&2&
-1&-2&
-2&4&-3 \\ \hline
%
\end{array}
$
\end{center}
\caption{\label{particle-content}\small The particle content of an $SO(10)$
  model with the family symmetry $PSL_2(7)$.}
\end{table}
Here $\psi$ denotes the ${\bf 16}$ of $SO(10)$ which incorporates the SM
fermions. Even though our model is based on $SO(10)$, it is convenient to
distinguish the left-handed and the right-handed components by showing
the decomposition into Pati-Salam representations. Using the order
$SU(4)\times SU(2)_L \times SU(2)_R$ we can write
\be
{\bf 16} ~\rightarrow~
\underbrace{({\bf 4},{\bf 2},{\bf 1})}_{\psi} ~+~
\underbrace{({\bf \ol 4},{\bf 1},{\bf 2})}_{\psi^c} \ .
\ee
$H_{10}$ and $H_{\ol{126}}$ are the $SO(10)$ Higgs fields whose $SU(2)_L$
doublet components enter the Yukawa couplings. In Pati-Salam language the
relevant components that acquire an electroweak VEV are
\be
H_{10} ~\rightarrow~ ({\bf 1},{\bf 2},{\bf 2}) \ , \qquad
H_{\ol{126}} ~\rightarrow~ ({\bf 15},{\bf 2},{\bf 2}) \  .
\label{Higgs}
\ee
The latter gives rise to the Georgi-Jarlskog (GJ) factor of 3 in the (2,2)
entry of the lepton mass matrix \cite{Georgi-Jarlskog,Review-inducedVEV}.
With the above specified fields, the leading Yukawa operators take the form,
\bea
\m L_{\mathrm{Yuk}} & \sim &
\frac{1}{M_H} \psi^{}_i   \hat\chi^{ij}_{\mathrm{top}} \psi_j^c H_{10}
~+~ \frac{1}{M^2} \psi^{}_i   \bar\phi^{i}_{23}  \bar\phi^{j}_{23} \psi_j^c
H_{\ol{126}}   \notag\\
&& + ~ \frac{1}{M^3} \psi^{}_i   (\bar\phi^{i}_{23}
\bar\phi^{j}_{123} + \bar\phi^{j}_{23}  \bar\phi^{i}_{123})\psi_j^c\,
\xi H_{10} \ ,
\eea
where we have only given the $PSL_2(7)$ indices $i,j$.
The corresponding diagrams are depicted in Fig.~\ref{yuk-fig}. Note that we
have inserted a $PSL_2(7)$ singlet flavon $\xi$ in order to additionally
suppress the third term.
Inserting the flavon VEVs, the first term ($a$) fills in the (3,3) entry
of the Yukawa matrix, the second ($b$) generates non-vanishing entries in the
2-3 block, and the third ($c$) enters everywhere in the Yukawa matrix except
for the (1,1) component. Including sextet and anti-triplet flavons, we find the following structure of the Yukawa matrices,
\bea
Y^{u,d,e,\nu} & \sim &
\frac{ |\vl \chi_{\mathrm{top}} \vr|   }{M_H}
\begin{pmatrix}
0 & 0 & 0 \\
0 & 0 & 0 \\
0 & 0 & 1
 \end{pmatrix} H_{10} ~+~
\frac{ |\vl \bar \phi_{23} \vr|^2   }{M^2}
 \begin{pmatrix}
0 & 0 & 0 \\
0 & 1 & -1 \\
0 & -1 & 1
 \end{pmatrix} H_{\ol{126}}
 \notag \\
&&\!\!\!+~
\frac{ |\vl \bar \phi_{23} \vr|\!\cdot \!|\vl \bar \phi_{123} \vr|  \,\vl\xi \vr}{ M^3}
\left[ \begin{pmatrix}
0 & 0 & 0 \\
1 & 1 & 1 \\
-1 & -1 & -1
 \end{pmatrix}
+
\begin{pmatrix}
0 & 1 & -1 \\
0 & 1 & -1 \\
0 & 1 & -1
\end{pmatrix}
\right]H_{10} \ .
\eea
To identify the Yukawa matrices for the different sectors $Y^{u,d,e,\nu}$
we note that the Higgs representations $H_{10}$ and $H_{\overline{126}}$
each have two Higgs doublets in them, according to Eq.~(\ref{Higgs}).
The two MSSM doublets $H_{u}$, $H_{d}$ originate below the GUT scale,
from one linear combination of these up-type doublets and one linear combination of the down-type ones
which remain almost massless. The orthogonal linear combinations
acquire GUT scale masses just as the colour triplets and other non-MSSM states.
Electroweak symmetry is broken after the light MSSM doublets $H_{u}$, $H_{d}$
acquire VEVs $v_{u,d}$ and they then generate the fermion masses. Since the dominant
contribution to the 2-3 block of the Yukawa matrices arises from
the components of $H_{u}$, $H_{d}$ coming from $H_{\overline{126}}$,
these entries will receive a relative Clebsch factor of 3 for the leptons
as compared to the quarks. Thus, ignoring messenger effects for the moment,
the resulting Dirac mass matrices take the form
\bea
m^{u,d,e,\nu} & \sim &
\begin{pmatrix}
0 & \epsilon^3 &  -\epsilon^3 \\
\epsilon^3 &  a \epsilon^2  &  - a \epsilon^2 \\
-\epsilon^3 & - a \epsilon^2   & 1
 \end{pmatrix} \frac{|\vl\chi_{\mathrm{top}} \vr|}{M_H} \, v_{u,d}  \ , ~\quad
\mathrm{with} ~~ \left\{ \begin{array}{ll} a=1 & \mathrm{for}~u,d \ , \\
    a=-3 & \mathrm{for}~e,\nu \ ,  \end{array} \right.~~ \label{dir-matrix}
\eea
where, for simplicity, we have chosen
\be
\frac{|\vl\chi_{\mathrm{top}} \vr|}{M_H} \sim 0.5 \ , \qquad
\frac{|\vl\bar\phi_{23} \vr|^2}{M^2}
\sim \epsilon^2 \ , \qquad
\frac{|\vl\bar\phi_{23}\vr|\!\cdot \!|\vl\bar\phi_{123} \vr| \, \vl\xi
  \vr}{M^3}  \sim \epsilon^3 \ .
\ee

\vspace{-0.5mm}

\noindent Notice the zero in the (1,1) entry whose presence allows to
accommodate the 
phenomenologically successful Gatto-Sartori-Tonin relation \cite{Gatto:1968ss}.
Moreover, the up and the down quark sectors can have independent messenger
masses $M\rightarrow M^u\!,M^d$, where $M^u\approx 3 M^d$ so that two different
expansion parameters $\epsilon \rightarrow \epsilon_u , \epsilon_d$ with
$\epsilon_u \approx \epsilon_d / 3$ are introduced as in
\cite{King:2003rf}. Numerically we need $\epsilon_u \approx 0.05$ and
$\epsilon_d \approx 0.15$. See \cite{Ross:2007az} for a detailed discussion of
the numerics including $\chi^2$ fits.  Thus the following mass matrix
structures for the quarks are quite elegantly achieved:
\be
m^{u}  \sim
\begin{pmatrix}
0 & \epsilon_u^3 &  -\epsilon_u^3 \\
\epsilon_u^3 &  \epsilon_u^2  &  -  \epsilon_u^2 \\
-\epsilon_u^3 & -  \epsilon_u^2   & 1
 \end{pmatrix} \frac{|\vl\chi_{\mathrm{top}} \vr|}{M_H} v_{u} \ , \ \ \ \
 m^{d}  \sim
\begin{pmatrix}
0 & \epsilon_d^3 &  -\epsilon_d^3 \\
\epsilon_d^3 &  \epsilon_d^2  &  -  \epsilon_d^2 \\
-\epsilon_d^3 & -  \epsilon_d^2   & 1
 \end{pmatrix} \frac{|\vl\chi_{\mathrm{top}} \vr|}{M_H} v_{d}\ ,
 \label{dir-matrix2}
\ee
with similar considerations in the lepton sector leading to $m^{e,\nu}$ being
the same as $m^{d,u}$ apart from the GJ factors of 3 in 
the 2-3 block, as in Eq.~(\ref{dir-matrix}). The (left-handed) quark mixing
angles in the up and the down sector are then calculated as
$\theta^{u,d}_{12}\sim \epsilon_{u,d}$, $\theta^{u,d}_{23}\sim
\epsilon^2_{u,d}$, and $\theta^{u,d}_{13}\sim \epsilon^3_{u,d}$, so that the
down sector gives the dominant contribution to a viable CKM matrix. Regarding
the leptons, the mixing angles are changed by GJ factors of 3 as
follows $\theta^{e,\nu}_{12}\sim \epsilon_{d,u}/3$, $\theta^{e,\nu}_{23}\sim 3
\epsilon^2_{d,u}$, and $\theta^{e,\nu}_{13}\sim \epsilon^3_{d,u}$. The
effect of the charged lepton mixing angles on the PMNS matrix is small so
that tri-bimaximal mixing is only perturbed within the experimentally allowed
range \cite{Antusch:2008yc}.

In this model the successful mass matrices in Eq.~(\ref{dir-matrix2})
are achieved in a very natural way
since, with the inclusion of the singlet flavon $\xi$, the first row and column
is cubic in the messenger mass, while the 2-3 block is quadratic and the 3-3
element involves the Higgs messenger mass $M_H$ and so is universal.

Turning to the Majorana sector different components of the field
$\Delta_{\ol{126}}$ are responsible for the masses of the left-handed and the
right-handed neutrinos. Analogous to the convention used to distinguish
the left-handed from the right-handed doublet components of $\psi$, we
introduce a notation that allows us to tell apart the left-handed and the
right-handed triplet components of $\Delta_{\ol{126}}$, i.e.
\be
{\bf \ol{126}} ~\rightarrow~
\underbrace{({\bf \ol{10}},{\bf 3},{\bf 1})}_{\Delta^{}_{\ol{126}}} ~+~
\underbrace{({\bf {10}},{\bf 1},{\bf 3})}_{\Delta^c_{\ol{126}}} ~+~ \cdots \ ,\label{decom-126bar}
\ee

\begin{figure}
\begin{center}
\includegraphics[width=9cm]{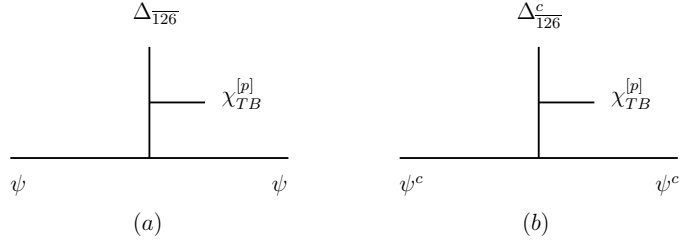}
\end{center}
\caption{\label{maj-fig}\small The diagrams which generate the Majorana
  couplings. In Pati-Salam language, $SU(4)\times SU(2)_L \times SU(2)_R$, the
  relevant components of the $SO(10)$ representations are:
  $(a)$ $\psi \rightarrow ({\bf 4},{\bf 2},{\bf 1})$,
  $\Delta_{\ol{126}} \rightarrow ({\bf \ol{10}},{\bf 3},{\bf 1})$ as well as
  $(b)$ $\psi^c \rightarrow ({\bf \ol 4},{\bf 1},{\bf 2})$,
  $\Delta^c_{\ol{126}} \rightarrow ({\bf {10}},{\bf 1},{\bf 3})$.}
\end{figure}

\vspace{-1mm}

\noindent where the ellipsis denotes the rest of the decomposition of the ${\bf
\ol{126}}$ of $SO(10)$ into representations of $SU(4)\times SU(2)_L \times
SU(2)_R$. With these remarks, the leading Majorana operators take the form
\bea
\m L_{\mathrm{Maj}} & \sim & \frac{1}{M}\sum_{p=0}^2 \left(
 \psi^{}_i
{\hat\chi^{[p] \, ij}_{TB}} \psi_j^{} \Delta_{\ol{126}}
~+~  \psi^{c}_i   \hat \chi^{[p]\,ij}_{TB}\psi_j^{c}
\Delta^c_{\ol{126}}  \right)  \ , \label{Maj-terms}
\eea
with the corresponding diagrams given in Fig.~\ref{maj-fig}.


While $\Delta^c_{\ol{126}}$ gets a GUT scale VEV in the triplet component of
$SU(2)_R$, $\Delta_{\ol{126}}$ acquires a small induced VEV
\cite{Review-inducedVEV}
\be \langle
\Delta_{\ol{126}} \rangle ~\sim~ \frac{v_u^2}{M} \ ,
\ee
in the triplet component of $SU(2)_L$,
for details see Appendix~\ref{app-induced}.
The first term in Fig.~\ref{maj-fig}\,($a$) then effectively generates a type II
see-saw contribution to the physical neutrino mass when the $\chi^{[p]}_{TB}$
and the $SU(2)_L$ triplet component of $\Delta_{\ol{126}}$ get their VEVs.
The resulting light neutrino mass matrix reads
\be
 m^\nu_{\mathrm{type\:II}} ~\sim ~  \sum_{p=0}^2
\vl \hat\chi^{[p]}_{TB} \vr  \cdot \frac{v_u^2}{M^2} \ , \label{type-II}
\ee
with the flavour structure encoded in the three matrices $\vl \hat
\chi^{[p]}_{TB} \vr$, $p=0,1,2$.

On the other hand, the second term of Eq.~(\ref{Maj-terms}) which is depicted
in Fig.~\ref{maj-fig}\,($b$) gives rise to the mass of the right-handed
neutrinos if $\Delta^c_{\ol{126}}$ acquires a VEV in the direction of the
right-handed triplet. In addition, the flavon
$\chi^{[p]}_{TB}$  which is required for $PSL_2(7) \times U(1)$ invariance
also needs to get a VEV. The existence of heavy right-handed neutrinos
unavoidably leads to a type I see-saw contribution to the light neutrino
masses. With the neutrino Dirac mass matrix
$m^\nu_{\mathrm{D}}$ of Eq.~(\ref{dir-matrix}), we find
\be
m^\nu_{\mathrm{type\:I}} ~\sim~ -\,
m^\nu_{\mathrm{D}}  \cdot
\left(
\sum_{p=0}^2  \frac{\vl \hat\chi^{[p]}_{TB} \vr \,   \vl
  \Delta^{c}_{\ol{126}} \vr}{M^2}
\right)^{-1}   \cdot ~
\frac{{m^\nu_{\mathrm{D}} }^T}{M} \ . \label{type-I}
\ee
The resulting effective light neutrino mass matrix then takes the form
\be
m^\nu_{\mathrm{eff}} ~=~ m^\nu_{\mathrm{type\:I}}~+~ m^\nu_{\mathrm{type\:II}}
\ .
\ee
In our $PSL_2(7)$ model of flavour we want the type~II contribution to
dominate over type~I. It is clear from Eqs.~(\ref{type-II},\ref{type-I}) how
this can be achieved. With $\vl  \chi^{[p]}_{TB} \vr
\sim  \vl   \Delta^{c}_{\ol{126}} \vr \sim M$ both contributions would be more
or less equally important. Increasing the scale of $\vl  \chi^{[p]}_{TB} \vr$
slightly, the type~II see-saw contribution increases while, at the same time,
the type~I see-saw gets suppressed. From now on we will assume dominance of
the type~II over the type~I see-saw.
The type~II see-saw mechanism has the additional advantage of avoiding the use
of operators with Clebsch zeros in the neutrino direction, a requirement
which is known to be essential in the type~I see-saw mechanism as applied to
these models in order to provide the necessary suppression required for TB
mixing\cite{King:2003rf}.

\section{Type II neutrino phenomenology}

In order to extract the neutrino mass spectrum that arises from the first term
of Eq.~(\ref{Maj-terms}), we need to insert the VEVs of the flavon sextets
$\chi^{[p]}_{TB}$, $p=0,1,2$ given in Eqs.~(\ref{tb-0-al}-\ref{tb-2-al}) into
Eq.~(\ref{yukmatrixreal}). Each of the three flavons comes with its own
coupling coefficient $c^{[p]}$ so that we have to diagonalise
\be
m^\nu_{\mathrm{eff}}~=~\frac{v_u^2}{M^2} \cdot
\sum_{p=0}^2 c^{[p]} \, \vl \hat \chi^{[p]}_{TB} \vr \ .
\ee
As each term is of tri-bimaximal form, we can diagonalise them individually
using the tri-bimaximal mixing matrix
\be
U_{TB}~=~ \begin{pmatrix}
\sqrt{\frac{2}{3}} & \sqrt{\frac{1}{3}} & 0 \\
-\sqrt{\frac{1}{6}} & \sqrt{\frac{1}{3}} & -\sqrt{\frac{1}{2}} \\
-\sqrt{\frac{1}{6}} & \sqrt{\frac{1}{3}} & \sqrt{\frac{1}{2}}
\end{pmatrix} \ .
\ee
This leads to\footnote{We implicitly absorb the potentially non-zero overall
  phases of the matrices $\vl \hat \chi^{[p]}_{TB}  \vr$ into a redefinition of the couplings $c^{[p]}$.}
\bea
m^\nu_{\mathrm{diag}} &=&
U_{TB}^{T} \, m^\nu_{\mathrm{eff}} \,  U_{TB}
~=~ \frac{(i-1)\bar b_7}{2\sqrt{3}} \cdot
 \frac{v_u^2}{M^2} \times \notag \\
&&  \times \left\{  \, c^{[0]} \, {|\vl \chi^{[0]}_{TB}
  \vr|}
  \cdot \mathrm{Diag}(1,1,-1)  \right. \notag \\
&&~+\: c^{[1]} \, {|\vl \chi^{[1]}_{TB}
  \vr|}
  \cdot
  \mathrm{Diag}\left(1,e^{i \varphi},-e^{-i \varphi}\right)
\notag \\
&&\left.~+\, c^{[2]} \, {|\vl \chi^{[2]}_{TB}
  \vr|}
  \cdot
  \mathrm{Diag}\left(e^{-i \varphi},e^{i \varphi},-1 \right)
\right\} \ ,
\eea
where the phase factor $e^{i \varphi}$ is fixed by the  $PSL_2(7)$ group
specific parameters $b_7$ and $\bar b_7$ as
\be
e^{i \varphi}~=~ \frac{\bar b_7}{b_7} ~=~\frac{(-3 + i \sqrt{7})}{4} \ ,
\ee
numerically corresponding to a phase of $\varphi \approx 138.6^\circ$. Then,
after dropping a global phase, the three (complex) light neutrino masses $m_i$
are each calculated as the sum of three mass parameters
\be
m^{[p]}~=~ \frac{1}{\sqrt{3}} \cdot
 \frac{v_u^2}{M^2} \cdot
  c^{[p]} \, {|\vl \chi^{[p]}_{TB}
  \vr|} \ ,
\ee
using different combinations of relative phase factors
\bea
m_1 &=& \phantom{-} m^{[0]} ~+~ \phantom{e^{-i \varphi}} m^{[1]}
                            ~+~ e^{-i \varphi} m^{[2]}  \ ,\notag \\
m_2 &=& \phantom{-} m^{[0]} ~+~ \;\;e^{i \varphi} m^{[1]}
                            ~+~ \;\;e^{i \varphi} m^{[2]}  \
                            , \label{nu-masses}\\
m_3 &=& - m^{[0]} ~-~ e^{-i \varphi} m^{[1]}
                  ~-~ \phantom{e^{-i \varphi}} m^{[2]}  \ .\notag
\eea

\begin{figure}
\begin{center}
\includegraphics[width=7cm]{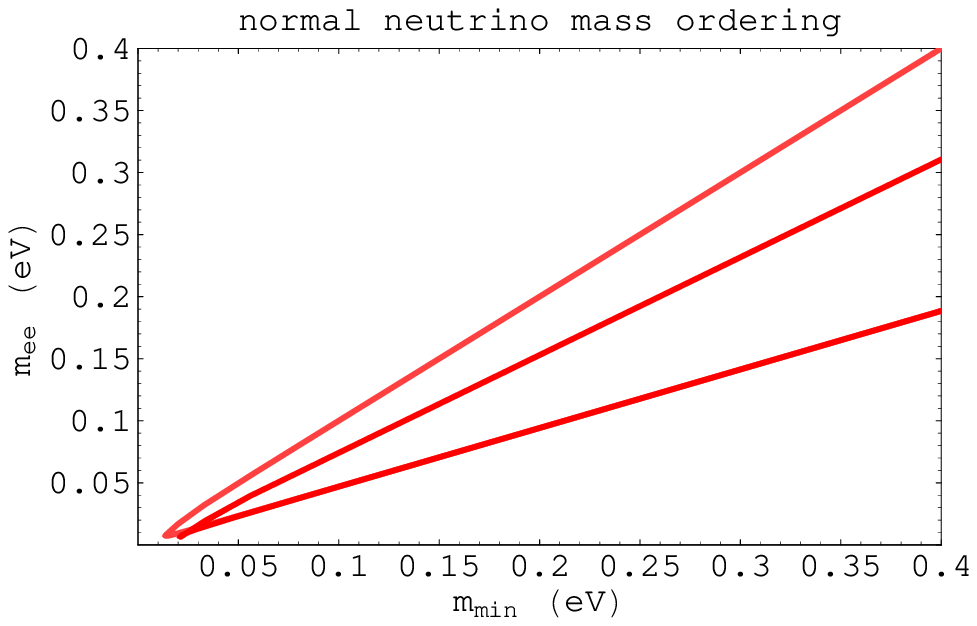} \qquad
\includegraphics[width=7cm]{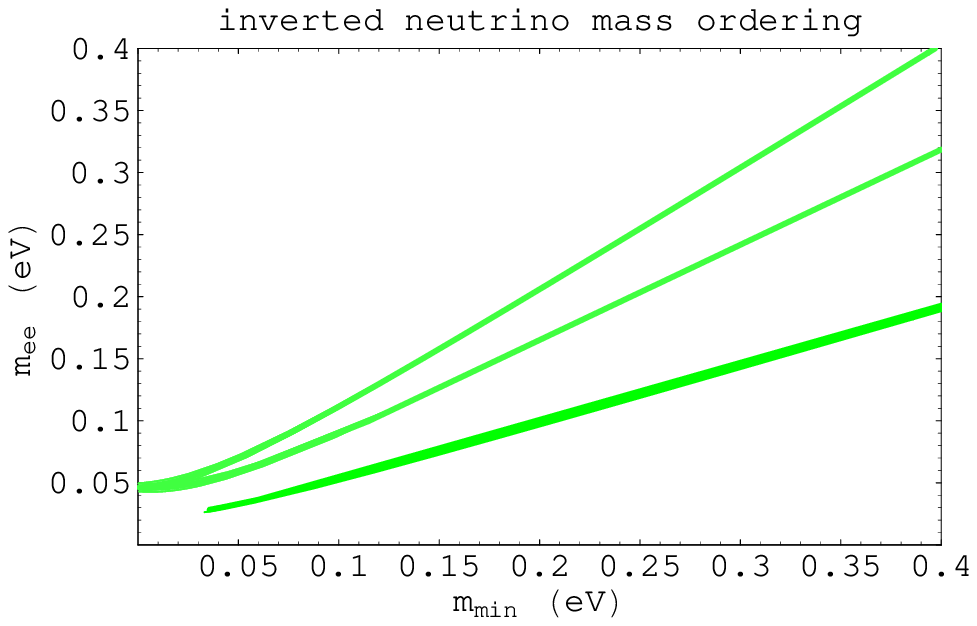}
\end{center}
\caption{\label{fig-Mee-Mmin}\small $m_{ee}$ plotted against $m_{\mathrm{min}}$ for the case of a normal mass ordering (left panel, in red) and an inverted
mass ordering (right panel, in green) assuming for simplicity real parameters $m^{[p]}$.}
\end{figure}

In the special case where the parameters $m^{[p]}$ are all real, it is
possible to find a very simple relation with the solar and the atmospheric
neutrino mass squared differences,
\bea
\Delta m_{\mathrm{sol}}^2 \,\:  &\equiv& |m_2|^2-|m_1|^2
~=~ \frac{7}{2} \cdot  m^{[1]} (m^{[2]}-m^{[0]}) \ ,\label{m0m1m2-s}\\
\Delta m_{\mathrm{atm}}^2  &\equiv&  |m_3|^2-|m_1|^2
~=~\frac{7}{2} \cdot   m^{[0]} (m^{[2]}-m^{[1]}) \ .\label{m0m1m2-a}
\eea
As $\Delta m_{\mathrm{sol}}^2$ and  $|\Delta m_{\mathrm{atm}}^2|$ have been
measured within certain error bars, it is straightforward to express
$m^{[1]}$ and $m^{[2]}$ as functions of $m^{[0]}$. Note that the sign
  ambiguity in $\Delta m_{\mathrm{atm}}^2$ leads to a case distinction between
  normal and inverted neutrino mass ordering.
With the (complex) masses $m_i$ taken from Eq.~(\ref{nu-masses}) we can
calculate the effective neutrino mass relevant for neutrinoless double beta
decay,
\be
m_{ee} ~=~ \Big| \sum_i  ({U_{TB}}_{ei})^2 m_i  \Big|
~=~  \Big| \frac{2}{3}\, m_1 + \frac{1}{3} \, m_2   \Big| \ .
\ee
In the case of real $m^{[p]}$ they will depend on $m^{[0]}$ only.
Furthermore we can easily determine the smallest (real-valued)
neutrino mass
\be
m_{\mathrm{min}} = \left\{
\begin{array}{ll}|m_1| \ , &  \mathrm{for~normal\ ,}\\
|m_3| \ , &  \mathrm{for~inverted\ ,}
\end{array} \right.
\ee
neutrino mass ordering which, in the case of
real $m^{[p]}$, is again only a function of $m^{[0]}$.
We can thus plot $m_{ee}$ against $m_{\mathrm{min}}$, as shown in Fig.~\ref{fig-Mee-Mmin}.

\begin{figure}
\begin{center}
\includegraphics[width=7cm]{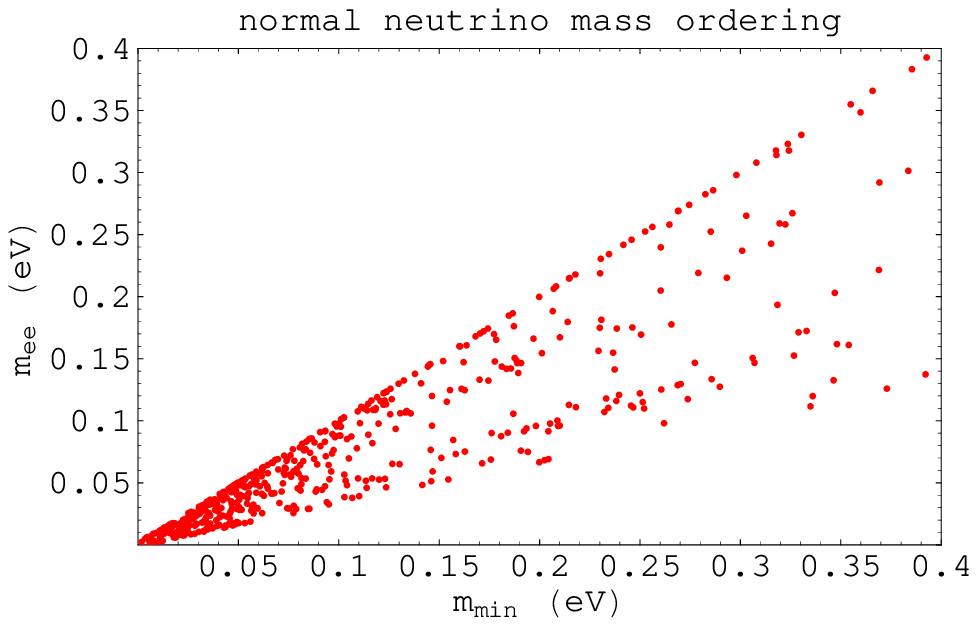} \qquad
\includegraphics[width=7cm]{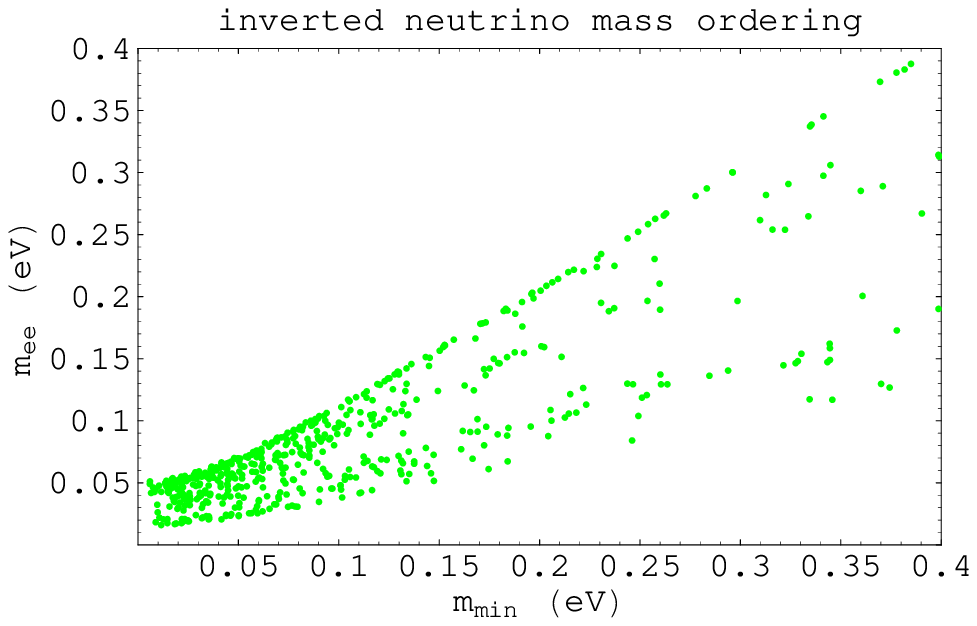}  \\[3mm]

\includegraphics[width=7cm]{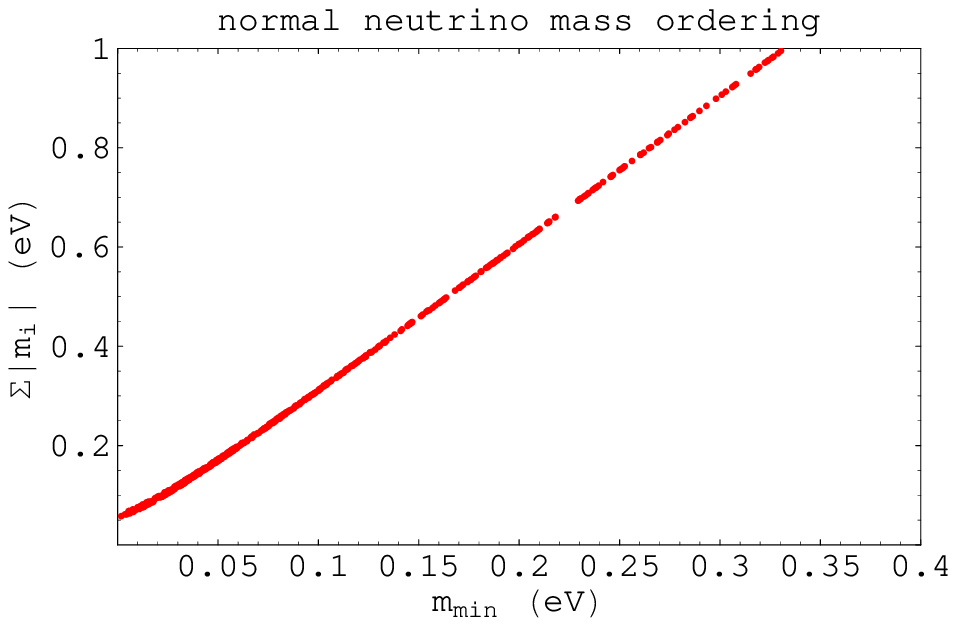} \qquad
\includegraphics[width=7cm]{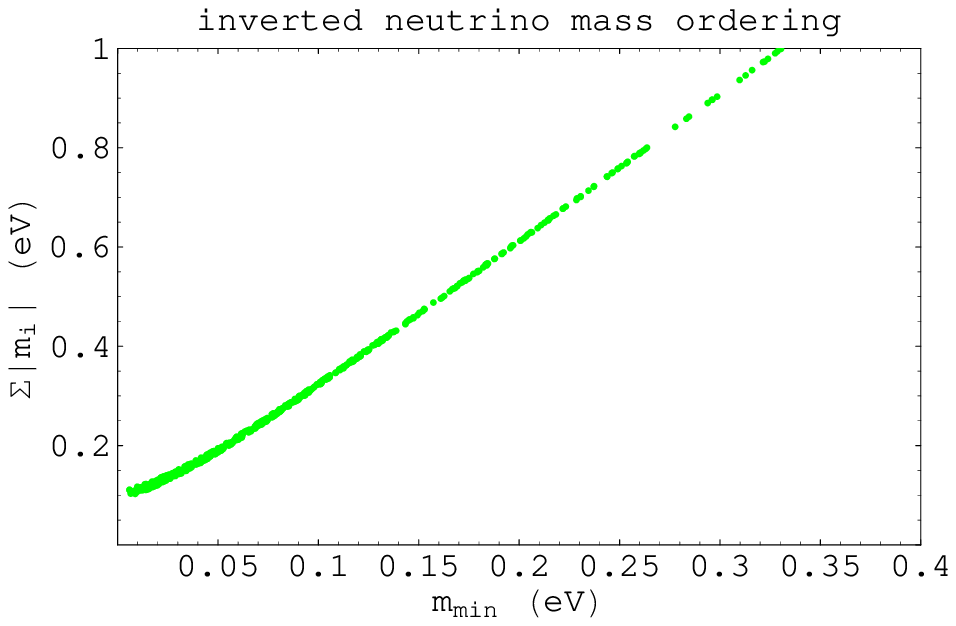}
\end{center}
\caption{\label{fig-Mee-Mmin-scatterLIN}
\small Scatter plots for $m_{ee}$ (upper panels) and $\sum_i
  |m_i|$  (lower panels) against $m_{\mathrm{min}}$ for the realistic case of complex parameters
$m^{[p]}$ for the case of a normal mass ordering (left panels, in red) and an inverted
mass ordering (right panels, in green).}
\end{figure}

Switching on the complex phases of the parameters $m^{[p]}$, the
dependence of $m_{ee}$ on $m_{\mathrm{min}}$ becomes much more fuzzy.
However, in order to illustrate the phenomenological consequences arising from
the general structure of Eq.~(\ref{nu-masses}) we have performed a scan
over the parameters $m^{[p]}=|m^{[p]}| e^{i\varphi^{[p]}}$ which we
have taken to be within the interval $0 \, \mathrm{eV} \le |m^{[p]}| \le
0.5 \, \mathrm{eV}$ with arbitrary phases $\varphi^{[p]}$. We have used equal
distribution in both the absolute values of the masses as well as the phases.
With such randomly generated input parameters it is
straightforward to calculate the resulting complex masses $m_i$ which in turn
can be converted into atmospheric and solar mass squared differences.
Keeping only those sets of parameters $m^{[p]}$ which lie within the
$3\sigma$~intervals~\cite{Schwetz:2008er}
\bea
2.07 \times 10^{-3} \,\mathrm{eV}^2 ~\le ~
|\Delta m_{\mathrm{atm}}^2|  &=& \Big||m_3|^2-|m_1|^2\Big|       ~\le~
2.75 \times 10^{-3} \,\mathrm{eV}^2 \ , \\
7.05 \times 10^{-5} \,\mathrm{eV}^2 ~\le ~\:
\phantom{|}\Delta m_{\mathrm{sol}}^2\phantom{|} \,  &=& \phantom{\Big|}|m_2|^2-|m_1|^2\phantom{\Big|}       ~\le~
8.34 \times 10^{-5} \,\mathrm{eV}^2 \ ,
\eea
we have calculated $m_{ee}$ and $m_{\mathrm{min}}$ in order to generate the
scatter plot version of Fig.~\ref{fig-Mee-Mmin}. The result is shown in
Fig.~\ref{fig-Mee-Mmin-scatterLIN}. We see that allowing for complex phase
factors fills in the gaps between the branches depicted in
Fig.~\ref{fig-Mee-Mmin}. In addition we also show the dependence
on $m_{\mathrm{min}}$ of
the sum of all light neutrino masses $\sum_i |m_i|$ which is relevant for
cosmology, where the current cosmological limit is about
$ \sum_i |m_i| < 1.0~\mathrm{eV}$ \cite{Lesgourgues:2006nd}.

\begin{figure}[t]
\begin{center}
\hspace{4mm}\includegraphics[width=7cm]{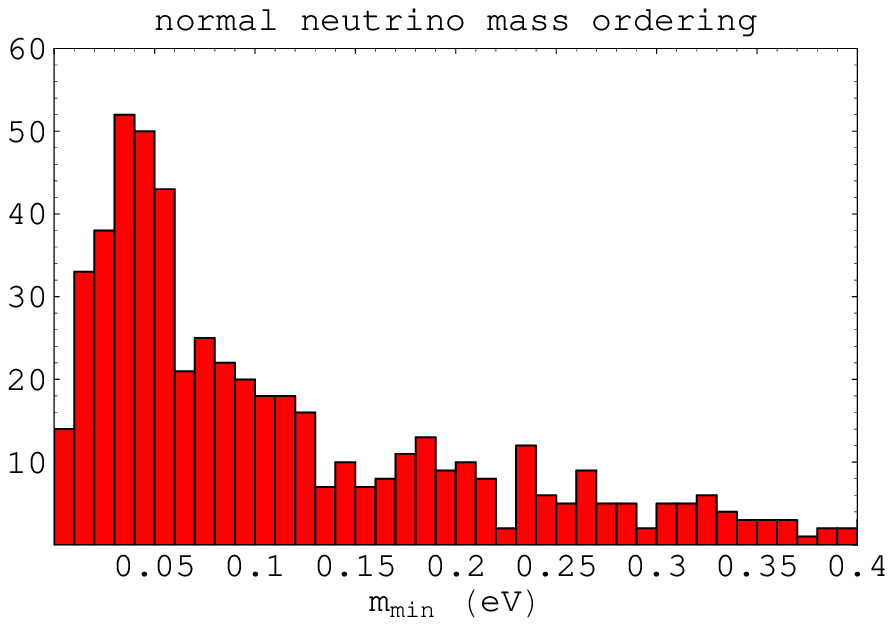}
\qquad
\includegraphics[width=7cm]{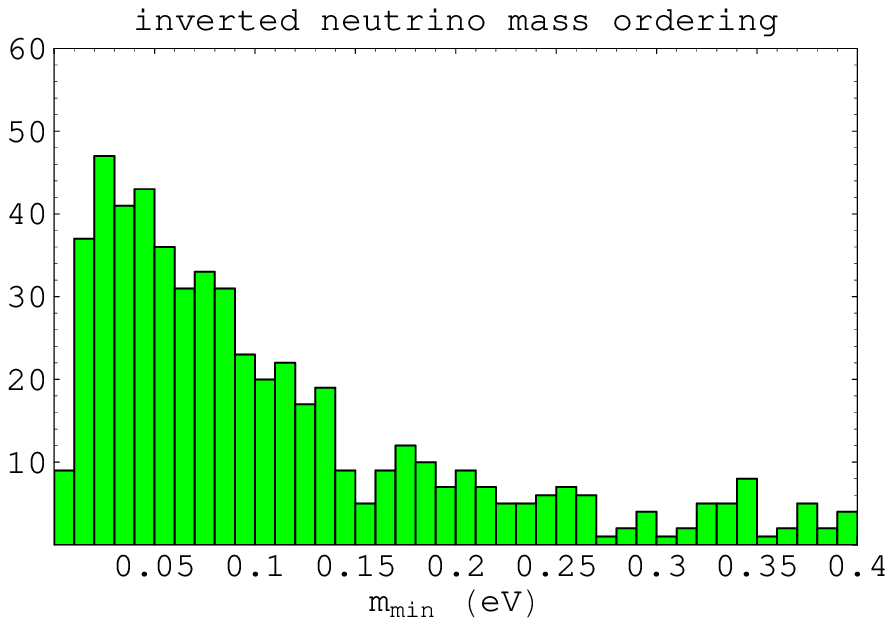} \\[3mm]
\hspace{4mm}\includegraphics[width=7cm]{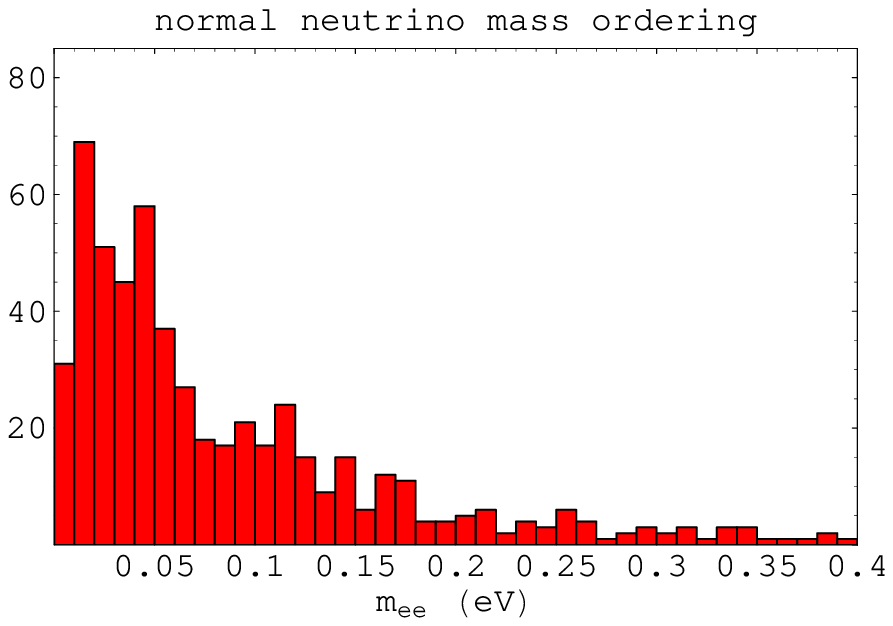}
\qquad
\includegraphics[width=7cm]{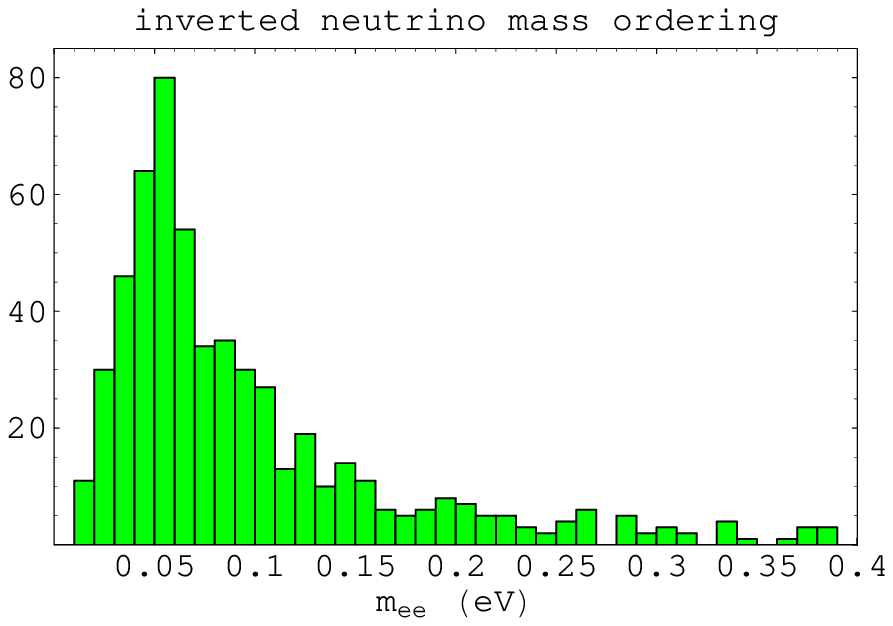}
\end{center}
\caption{\small \label{fig-histograms-Mee-Mmin-scatterLIN}These histograms show the distribution of the numbers of points
as a function of $m_{\mathrm{min}}$ (upper panels) and as a function of
$m_{ee}$ (lower panels) corresponding to the scatter plots in
Fig.~\ref{fig-Mee-Mmin-scatterLIN}, for the case of
a normal mass ordering (left panels, in red) and an inverted
mass ordering (right panels, in green). These histograms all
exhibit a very broad peak at about 0.05~eV with significant tails out to
about 0.4 eV.}
\end{figure}

The histograms in Fig.~\ref{fig-histograms-Mee-Mmin-scatterLIN}
show the distribution of the numbers of points
as a function of $m_{\mathrm{min}}$ (upper panels) and as a function of
$m_{ee}$ (lower panels) corresponding to the scatter plots in
Fig.~\ref{fig-Mee-Mmin-scatterLIN}, for the case of
a normal mass ordering (left panels, in red) and an inverted
mass ordering (right panels, in green).
All the distributions
exhibit a very broad peak at about $\sim 0.05~\mathrm{eV}$ with significant tails out to about $\sim 0.4~\mathrm{eV}$ in both $m_{\mathrm{min}}$ and $m_{ee}$.

\section{\label{vac-sec}Vacuum Alignment in the $\bs{PSL_2(7)\times SO(10)}$ model}
\cleqn
We now return to the question of vacuum alignment in the
$PSL_2(7)\times SO(10)$ model, using the  $D$-term method
and starting with the alignment of the flavon sextets.
It turns out that, for $PSL_2(7)$ triplets and anti-triplets,
the $D$-term vacuum alignment method reviewed recently in \cite{King:2009ap}
does not work since the invariants involving triplets are the same as in $SU(3)$
and the new invariants required for triplet flavon vacuum alignment are not
present. However, as we shall discuss below, there are new invariants
involving sextets so the $D$-term method of vacuum alignment is well suited to
aligning the flavon sextets.
Once the sextet flavons are properly aligned, the anti-triplet flavons
may then be aligned by coupling them to the sextet flavons, as we shall also
show. This implies that $PSL_2(7)$ sextets play the crucial role in vacuum
alignment quite independently of the crucial role that they play in generating
third family Yukawa couplings and Majorana masses.

\subsection{Invariants with sextets}

We determine the quartic invariants for the sextet flavon $\chi$. Although the
sextet irrep of $PSL_2(7)$ is real, the physical field need {\it not} be
real. In general we therefore consider invariants of type
$\chi^{\dagger}_i\chi_j^{}\chi^{\dagger}_k\chi_l^{}$,
where we shall assume that such terms involve only
sextet flavons of a particular type, i.e. only purely
$\chi^{[p]}_{TB}$ (for a particular choice of $p$) or
the flavon $\chi^{}_{\mathrm{top}}$ but not operators involving
a mixture of different flavon types.
This may be enforced by assuming
a particular messenger sector and introducing a
$U(1)'$ symmetry as discussed in Appendix~\ref{app-control}.
Since we are interested in operators involving only one flavon type, we have to
consider only the {\it symmetric} combinations
$$
({\bf 6 \otimes 6})_s
~=~
{\bf 1 ~\oplus~ 6 ~\oplus~ 6 ~\oplus~ 8} \ .
$$
For $\chi$ given in the real sextet basis of \cite{King:2009mk}, the
representations on the right-hand side can be written as

\bea
{\bf 1}:~\quad \Upsilon &=& \sum_{i=1}^6 \chi_i \chi_i \ , \\[4mm]
{\bf 6}:~\quad \Theta &=&
\begin{pmatrix}
-2\sqrt{21}\chi_1\chi_5 - 6\chi_1\chi_6 \\[1mm]
2\sqrt{21}\chi_2\chi_5 - 6\chi_2\chi_6 \\[1mm]
2\sqrt{14}\chi_3\chi_4 + 8 \chi_3\chi_6 \\[1mm]
\sqrt{14} \chi_3^2 - \sqrt{14} \chi_4^2 + 8 \chi_4 \chi_6 \\[1mm]
-\sqrt{21} \chi_1^2 +\sqrt{21} \chi_2^2 - 6 \chi_5\chi_6 \\[1mm]
-3 \chi_1^2 -3 \chi_2^2 + 4 \chi_3^2 +4  \chi_4^2 -3 \chi_5^2 + \chi_6^2
\end{pmatrix} \ , \label{eq-6sym}\\[4mm]
{\bf 6}:~\quad \Theta' &=&
\begin{pmatrix}
2\sqrt{21}\chi_2\chi_3 + 2\sqrt{7}\chi_1\chi_4 - 2 \sqrt{2}\chi_1\chi_6\\[1mm]
2\sqrt{21}\chi_1\chi_3 + 2\sqrt{7}\chi_2\chi_4 - 2 \sqrt{2}\chi_2\chi_6\\[1mm]
2\sqrt{21}\chi_1\chi_2 + 2\sqrt{7}\chi_3\chi_4 - 2 \sqrt{2}\chi_3\chi_6\\[1mm]
\sqrt{7}\chi_1^2 +  \sqrt{7}\chi_2^2 + \sqrt{7}\chi_3^2 - \sqrt{7}\chi_4^2
-2 \sqrt{7}\chi_5^2 - 2  \sqrt{2}\chi_4\chi_6 \\[1mm]
-4\sqrt{7}\chi_4\chi_5 - 2 \sqrt{2}\chi_5\chi_6 \\[1mm]
-\sqrt{2}\chi_1^2 - \sqrt{2}\chi_2^2 - \sqrt{2}\chi_3^2 - \sqrt{2}\chi_4^2 -
\sqrt{2}\chi_5^2 + 5 \sqrt{2}\chi_6^2
\end{pmatrix}\ , \label{eq-6symprime}  \\[4mm]
{\bf 8}:~\quad \Omega &=&
\begin{pmatrix}
\sqrt{3} \chi_2\chi_3 + \chi_1\chi_4 -2\sqrt{6} \chi_1\chi_5 +2\sqrt{14}
\chi_1\chi_6 \\[1mm]
 \sqrt{21}\chi_2\chi_3 -3\sqrt{7} \chi_1\chi_4 \\[1mm]
-\sqrt{6}\chi_1^2 +\sqrt{6}\chi_2^2 -2\chi_4\chi_5 +2\sqrt{14}\chi_5\chi_6
\\[1mm]
- \sqrt{2}\chi_1^2  -  \sqrt{2}\chi_2^2
+2\sqrt{2}\chi_3^2-2\sqrt{2}\chi_4^2+2\sqrt{2}\chi_5^2-2
\sqrt{7}\chi_4\chi_6  \\[1mm]
-\sqrt{21}\chi_1\chi_3  +3\sqrt{7} \chi_2\chi_4\\[1mm]
\sqrt{3}
\chi_1\chi_3+\chi_2\chi_4+2\sqrt{6}\chi_2\chi_5+2\sqrt{14}\chi_2\chi_6\\[1mm]
-2\sqrt{21}\chi_3\chi_5\\[1mm]
2\sqrt{6}\chi_1\chi_2-4\sqrt{2}\chi_3\chi_4+2\sqrt{7}\chi_3\chi_6
\end{pmatrix}\ .\label{Delta}
\eea
It should be mentioned that the two sextets $\Theta$ and $\Theta'$ are not
defined uniquely since any linear combination of $\Theta$ and $\Theta'$
transforms as a ${\bf 6}$ as well. This ambiguity doesn't affect the {\it set}
of quartic invariants even though the explicit form of a particular invariant
may be different. For the octet $\Omega$ we have chosen a basis where the
generators $\m S^{[{\bf 8}]},\m T^{[{\bf 8}]},\m U^{[{\bf 8}]},\m V^{[{\bf 8}]}$
are real with $\m S^{[{\bf 8}]}$ and $\m U^{[{\bf 8}]}$ diagonal. They can be
found in Appendix~\ref{octet}. The derived quartic invariant is necessarily
independent of the choice for the octet basis.

The Kronecker product for the symmetric combinations $({\bf 6 \otimes 6})_s$
shows that there are six independent quartic invariants of type
$\chi^\dagger_i\chi^{}_j\chi^\dagger_k\chi^{}_l$. They can be easily obtained
from $\Upsilon$,~$\Theta$, $\Theta'$, $\Omega$ and their complex conjugates,
denoted by $\bar\Upsilon$, $\bar\Theta$, $\bar\Theta'$, $\bar\Omega$.
Since we have chosen real bases, the quartic invariants are formed trivially:
\be
\m I_1~=~ \bar\Upsilon \, \Upsilon \ , ~\qquad
\m I_2~=~ \sum_{i=1}^6 \bar\Theta_i  \,\Theta_i \ , ~\qquad
\m I_3~=~ \sum_{i=1}^6 \bar\Theta'_i  \,\Theta'_i \ ,
\ee
\be
\m I_4~=~  \frac{1}{\sqrt{2}}\sum_{i=1}^6 (\bar\Theta'_i  \,\Theta^{}_i +
\Theta'_i \, \bar\Theta^{}_i )\ , ~\qquad
\m I_5~=~  \frac{i}{\sqrt{2}}\sum_{i=1}^6 (\bar\Theta'_i  \,\Theta^{}_i -
\Theta'_i \, \bar\Theta^{}_i )\ ,
\ee
\be
\m I_6~=~ \sum_{a=1}^8 \bar\Omega_a  \,\Omega_a \ .
\ee
Here we have chosen a convention in which all invariants are real. The way we
constructed these six invariants obscures a trivial one, namely
$\left(\sum_i\chi^\dagger_i\chi^{}_i\right)^2$. It is related to the above
invariants by
\be
\m I_0~\equiv~ \left(\sum_i\chi^\dagger_i\chi^{}_i \right)^2 ~=~
\frac{1}{6 \cdot 49} \,
\left(49 \, \m I_1 + 5 \, \m I_2 + 5 \, \m I_3 - \m I_4 + 7 \, \m I_6 \right)
\ .
\ee
It is therefore possible to replace $\m I_6$ in favour of $\m I_0$ in our set
of independent quartic invariants.\footnote{In $SU(3)$ the product $({\bf
    6\otimes 6})_s \otimes ({\bf \ol 6\otimes \ol 6})_s$ yields only two
  independent invariants: $\m I_0$ and the sum $2\, \m I_2+2\,\m I_3 +\m
  I_4-\sqrt{7}\,\m I_5$.}

\subsection{A potential for obtaining sextet flavon alignments}

Let us study a sextet potential of the form \cite{King:2009ap}
\bea
V &=& -m^2 \cdot \sqrt{\m I_0}
~+~ \lambda \cdot \m I_0
~+~ \lambda \cdot \sum_{\alpha=1}^5 \kappa_\alpha \, \m I_\alpha  \notag \\
&=&-m^2 \cdot \sqrt{\m I_0}
~+~ (\lambda \cdot \m I_0 ) \cdot f
\ , \label{Vpot}
\eea
where we have defined
\be
f\equiv \left( 1 ~+~  \sum_{\alpha=1}^5 \kappa_\alpha \,
  \frac{\m I_\alpha}{\m I_0} \right).
\ee
In order for this potential to have a minimum, $\lambda \cdot f$ must be
positive. Moreover, the factor $f$ is independent of the normalisation of
$\langle \chi \rangle $ so that $f$ is extremised by an appropriate choice of
the alignment. On the other hand, $\m I_0$ is independent of a particular
alignment. Therefore, denoting the extremum of $f$ by $f_0$, the overall scale
of the minimum is determined from
$$
\sum_{i=1}^6 \vl \chi^\dagger_i\vr\vl \chi^{}_i\vr
~=~
\sqrt{\vl \m I_0 \vr} ~=~ \frac{m^2}{2 \, \lambda \cdot f_0} \ .
$$
In the following we will assume $\lambda>0$. So we need to find the {\it
  minimum} $f_0>0$ of~$f$. However, since we already know the desired alignment
vectors, our procedure will be to insert these into $f$ and find suitable
coefficients $\kappa_\alpha$ such that $f$ becomes a (local) minimum.

\subsubsection{${\bs{\chi_{TB}^{[p]}}}$ alignment}
Consider the alignment of Eq.~(\ref{tb-0-al}) which breaks $PSL_2(7)$ down to
$S_4$
\be
\langle \tilde \chi^{}_{\m{}} \rangle ~ \propto ~  (0\,,\,0\,,\,0\,,\,0\,,\,0
\,,\,1)^T \ . 
\label{minalign1}
\ee
We find vanishing first derivatives for each individual invariant,
$$
\frac{\partial (\m I_\alpha / \m I_0)}{\partial (\mathrm{Re}\, \chi_i)} ~=~
\frac{\partial (\m I_\alpha / \m I_0)}{\partial (\mathrm{Im}\, \chi_i)}~=~0 \
,  \quad \forall ~\alpha=1,...,5\ , ~~~\forall ~ i=1,...,6 \ .
$$
In order to get a minimum, the $12 \times 12$ matrix of second derivatives
(the Hessian $H_\alpha$) needs to be positive-definite except for those two real
directions which give zero since they correspond to the invariance of $f$
under the scaling $\chi_i ~\rightarrow ~ s \cdot \chi_i$, with $s\in
\mathbb{C}$. With the alignment $\langle \tilde \chi^{}_{\m{}}\rangle$, this
is the $\chi_6$
direction. For the remaining 10 real dimensions, we obtain the following
matrices $h_\alpha$
\bea
h_1 &=&
-8 \cdot
\text{Diag}\,(0\,,\,0\,,\,0\,,\,0\,,\,0\,,\,1\,,\,1\,,\,1\,,\,1\,,\,1) \ ,\\[1mm]
h_2 &=& 4\cdot
\text{Diag}\,(14\,,\,14\,,\,35\,,\,35\,,\,14\,,\,20\,,\,20\,,\,27\,,\,27\,,\,20)\ ,\\[1mm]
h_3 &=& -16 \cdot
\text{Diag}\,(14\,,\,14\,,\,14\,,\,14\,,\,14\,,\,9\,,\,9\,,\,9\,,\,9\,,\,9)\ , \\[1mm]
h_4 &=& -4\cdot
\text{Diag}\,(14\,,\,14\,,\,7\,,\,7\,,\,14\,,\,-18\,,\,-18\,,\,45\,,\,45\,,\,-18) \ ,\\[1mm]
h_5 &=& \begin{pmatrix} 0 & \tilde h_5 \\ \tilde h_5 & 0 \end{pmatrix} \ ,
\qquad
\tilde h_5 ~=~ 28 \cdot \text{Diag}\,(2\,,\,2\,,\,-3\,,\,-3\,,\,2)  \ .
\eea
The condition for a minimum of the potential is that all the 
eigenvalues of the $10 \times 10 $
matrix
$$
h~=~\sum_{\alpha=1}^5 \kappa_\alpha \, h_\alpha \ ,
$$
should be positive-definite for appropriate $\kappa_\alpha$. The explicit forms of
$h_\alpha$ show that many potentials can be constructed that are minimised by
the  alignment of Eq.~(\ref{minalign1}). For instance, one could choose
$$
-2~<~\kappa_1~<~10 \ , \qquad \kappa_2~=~1 \ , \qquad \kappa_3~=~\kappa_4~=~\kappa_5 ~=~0\ .
$$
The upper bound on $\kappa_1$ arises because the sum $\kappa_1h_1+\kappa_2h_2$
must be positive-definite. The lower bound, on the other hand, is related to
the requirement of $f>0$. Another possibility
would be
$$
-\frac{1}{50}~<~\kappa_3~<~0 \ ,  \qquad
\kappa_1~=~\kappa_2~=~\kappa_4~=~\kappa_5 ~=~0\ ,
$$
where the lower bound arises due to $f>0$.
We emphasise here that both examples do not rely on any tuning and the 
vanishing of the $\kappa$ coefficients, as assumed in these examples, 
is not a requirement to get a positive-definite $h$.
It is not our intention to give the most general potential that is minimised
by the alignment $\langle \tilde \chi^{}_{\m{}}\rangle$, but merely to show
that a potential leading to such an alignment is quite plausible. These examples are sufficient to show this.

Due to the symmetry of such a potential under $PSL_2(7)$, there exists a
discrete degeneracy of minima. Given a potential $V$ that is minimised by
$\langle \tilde \chi^{}_{\m{}}\rangle$, also the alignment $\m G \cdot
\langle \tilde \chi^{}_{\m{}}\rangle$, with $\m G$ denoting any element of
$PSL_2(7)$ in the sextet representation, yields a minimum. In addition to
$\langle\tilde \chi^{}_{\m{}}\rangle\equiv\langle\tilde \chi^{[0]}_{\m{}}\rangle$,
we find six new alignment vectors:
\bea
\langle\tilde \chi^{[1]}_{\m{}}\rangle &\propto& \frac{1}{6} \cdot (0\,,\,
0\,,\,0 \,,\, -\sqrt{14}\,,\,-\sqrt{21} \,,\, -1)^T \ , \\[2mm]
\langle\tilde \chi^{[2]}_{\m{}}\rangle &\propto& \frac{1}{6} \cdot(0\,,\,
0\,,\,0 \,,\, -\sqrt{14}\,,\,\sqrt{21} \,,\, -1)^T \ ,\\[2mm]
\langle\tilde \chi^{[3]}_{\m{}}\rangle &\propto& \frac{1}{6\sqrt{2}}
\cdot(\sqrt{21}\,,\, \sqrt{21}\,,\,\sqrt{21} \,,\, \sqrt{7}\,,\, 0  \,,\,
-\sqrt{2})^T \ ,\\[2mm]
\langle\tilde \chi^{[4]}_{\m{}}\rangle &\propto&\frac{1}{6\sqrt{2}} \cdot
(\sqrt{21}\,,\, -\sqrt{21}\,,\,-\sqrt{21}\,,\, \sqrt{7}\,,\, 0  \,,\,
-\sqrt{2})^T \ ,\\[2mm]
\langle\tilde \chi^{[5]}_{\m{}}\rangle &\propto&\frac{1}{6\sqrt{2}} \cdot
(-\sqrt{21}\,,\, \sqrt{21}\,,\,-\sqrt{21} \,,\, \sqrt{7}\,,\, 0  \,,\,
-\sqrt{2})^T \ ,\\[2mm]
\langle\tilde \chi^{[6]}_{\m{}}\rangle &\propto&\frac{1}{6\sqrt{2}} \cdot
(-\sqrt{21}\,,\, -\sqrt{21}\,,\,\sqrt{21} \,,\, \sqrt{7}\,,\, 0  \,,\,
-\sqrt{2})^T \ .
\eea
The alignment vectors $ \langle \tilde \chi^{[p]}_{\m{}}\rangle$ with
$p=0,1,2$ have the form of Eq.~(\ref{TBalignment}) and correspond to
the vectors $\langle \chi^{[p]}_{TB}\rangle$ of
Eqs.~(\ref{tb-0-al}-\ref{tb-2-al}). Therefore, three of the seven discrete
minima  lead to a tri-bimaximal structure.

It is worth mentioning that the above vacuum alignments $\vl \tilde \chi^{[p]} \vr$ can alternatively also be obtained form an $F$-term alignment mechanism.
Consider a superpotential of the form
\be
W~=~ \m{M} \, \chi_0^{} \chi
~+~ g\, \chi_0^{} (\chi\chi)^{}_{\bf 6}
~+~ {g}^\prime \chi_0^{} (\chi\chi)^\prime_{\bf 6} \ ,
\ee
where $\chi^{}_0$ is a driving sextet field. The parentheses denote the
contraction to the two distinct sextet as shown in
Eqs.~(\ref{eq-6sym},\ref{eq-6symprime}). It is easy to show that the resulting
$F$-term equations of $\chi^{}_0$ are solved by the sextet alignment $\vl
\tilde\chi^{[0]}_{}\vr = (0,0,0,0,0,v)^T$ if
$$
v~=~-\,\frac{\m M}{g+5 \sqrt{2}g^\prime_{}} \ .
$$
Due to the symmetry of the superpotential under $PSL_2(7)$, the other
six alignments $\vl \tilde\chi^{[p]}_{}\vr$ lead to vanishing $F$-terms as
well. Thus the three sextets leading to a tri-bimaximal structure can be
aligned alternatively through an $F$-term mechanism. In the case of charged
sextets the mass parameter $\m M$ must be generated dynamically by the VEV of
some $PSL_2(7)$ singlet.

\subsubsection{${\bs{\chi^{}_{\mathrm{top}}}}$ alignment}
We now turn to the study of the alignment of
Eq.~(\ref{topalignment}). 
In general we seek to solve ten non-trivial conditions
with a set of five parameters $\kappa_i$. Therefore it is by no means guaranteed
that an extended parameter space of solutions to the minimisation conditions of the
potential should exist. Remarkably, by 
plugging Eq.~(\ref{topalignment}) into the potential in
Eq.~(\ref{Vpot}), we find that, for a certain choice of parameters discussed below,
the first derivatives of $f$ can be made to vanish. However, unlike the previous
alignments, first derivatives of $f$ do not vanish 
for all the invariants taken separately. 

To be precise, a straightforward calculation shows that,
in order to achieve the desired top alignment $\vl \chi^{}_{\mathrm{top}} \vr$, 
the vanishing of the first derivatives
requires {\it only two} relations amongst the different sextet combinations
which result from the symmetric combinations of two sextets, to wit
$\kappa_2=\kappa_3=\kappa_4+ \kappa_5 / \sqrt{7}$. Clearly the origin of these
relations remains to be understood, and, for example, could result from
some underlying higher symmetry, although this goes beyond the scope of
the present $PSL_2(7)$ discussion. However we emphasise that the fact that 
a successful potential can be found with particular
values of parameters which can lead to the desired top alignment
is highly non-trivial and this is not the case in general for other
alignments.\footnote{To illustrate this, consider the alignment $\vl
  \chi^{}_{\mathrm{alt}} \vr \propto (-1,0,0,\sqrt{3}/b7,\sqrt{2},0)^T$ which
  leads to a Yukawa matrix with identical $(2,2)$ and $(3,3)$ elements and
  zero entries everywhere else. Requiring vanishing first derivatives fixes
  the parameters $\kappa_i$ uniquely. Up to an overall scale we find
  $\kappa_1=0$, $\kappa_2=\kappa_3=2 \kappa_4 = 2 \kappa_5
  /\sqrt{7}$. However, in this case, the matrix of second derivatives turns
  out to have positive and negative eigenvalues showing that no quartic
  potential exists which is minimised by $\vl \chi^{}_{\mathrm{alt}} \vr$.}

Assuming the above two relations, the potential that has a chance of being
minimised by $\langle \chi^{}_{\mathrm{top}}\rangle$ includes an $f$ of the form
\bea
f &=& \frac{1}{\m I_0} \left[ \m I_0 ~+~ \kappa\,
\underbrace{\m I_1}_{\equiv \,\m I}
~+~ \kappa'
\underbrace{\left( \m I_2\,+\, \m I_3 \,+\, \m I_4 \right)}_{\equiv \,\m I'}
~+~ \kappa''
\underbrace{\left( \m I_4\,-\,\sqrt{7}\,\m I_5 \right)}_{\equiv \, \m I''}
\right] \ .
\eea
Since $f$ has to be positive, we get the condition
\be
0 ~<~ f\Big|_{\langle \chi^{}_{\mathrm{top}} \rangle}
~=~ 1~+~ 0 \cdot \kappa ~+~ \frac{70}{3} \cdot \kappa'
~+~ \frac{140}{3} \cdot \kappa'' \ .
\ee
On the other hand, the Hessian needs to be positive-definite. The calculation
for all three independent contributions to $f$ shows that the 10-dimensional
Hessian $h$ has two positive and eight zero eigenvalues, while $h'$ has six
positive and four negative ones, and finally $h''$ has ten negative
eigenvalues. We therefore conclude that the choice
\be
\kappa~=~\kappa'~=~0 \ ,\qquad  -\frac{3}{140} ~<~\kappa''~<~0 \ ,
\ee
leads to a potential $V$ which is minimised by the alignment of
Eq.~(\ref{topalignment}). In other words, $\m I''$ must be suppressed compared
to $\m I_0$. Other solutions are possible as well, but as the three Hessians
cannot be diagonalised simultaneously, it is not easy to give a general
expression. However we emphasise that $\kappa$ and $\kappa'$ need not be zero,
for instance the choice
$$
\kappa~=~\kappa'~=~-\kappa''~=~\frac{1}{100} \ ,
$$
leads to an acceptable potential as well.

Having illustrated that the alignment $\langle \chi^{}_{\mathrm{top}}\rangle$
can be obtained from reasonable potentials, we again find that a
transformation of this alignment vector under $PSL_2(7)$ yields new minima. In
this case, we get 168 different minima, including the one of
Eq.~(\ref{topalignment}). All of them are physically identical since we can use
the freedom to redefine our basis by applying a suitable $PSL_2(7)$ symmetry
transformation.

\subsection{Flavon anti-triplet vacuum alignment}
\cleqn
As outlined in \cite{King:2009mk} we need to introduce anti-triplet flavon
fields to generate the first and second family Yukawa couplings. The question
arises how to align such anti-triplets $\bar \phi$ of $PSL_2(7)$. The
immediate idea to study quartic terms  of the form ${\bar \phi_i}^\dagger {\bar
\phi^j}_{} {\bar \phi_k}^\dagger {\bar \phi^l}_{}$, which proves successful
for flavour groups like
$\Delta_{27}$ \cite{deMedeirosVarzielas:2006fc},
$Z_7\rtimes Z_3$ \cite{Z7Z3-LQ},
$A_4$ \cite{King:2006np} and
others \cite{King:2009ap}, does not lead to an alignment of the anti-triplets
because $PSL_2(7)$ is too big a symmetry so that the trivial $SU(3)$
invariant ${\bar \phi_i}^\dagger {\bar \phi^i}_{} {\bar \phi_j}^\dagger {\bar
\phi^j}_{}$ is the only allowed quartic term of only anti-triplets.

However, due to the presence of the sextet representation, the possibility of
aligning the anti-triplets by coupling them to the pre-aligned sextets
arises. It turns out that the simplest operator
$\bar \phi\, \chi^{[0]}_{TB} \,\bar \phi $ is not suitable.
After inserting the VEV of $\chi^{[0]}_{TB}$ we obtain a contribution to the
anti-triplet scalar potential which is extremised by the alignment
$\vl \bar \phi \vr  ~\propto ~ (1 , x , \bar x)$.
Due to the degeneracy of vacua the desired anti-triplet alignments of
Eq.~(\ref{anti-tri-ali}) cannot be generated form terms like $\bar \phi\,
\chi^{[p]}_{TB} \,\bar \phi $, where $p=0,...,6$.\footnote{Note, however, that
  it is possible to use similar couplings to align anti-triplets against the
  $T$   preserving sextet $\chi_T^{}$ with an alignment of the form  $\vl
  \chi_T^{}\vr \propto (\sqrt{2},0,0,0,1,z)^T$, where $z$ remains
  undetermined. It turns out that the resulting anti-triplet alignment $\vl
  \bar \phi \vr ~\propto ~ (1 , 0 , 0)$ preserves $T$ as well.}

Since the simplest operator discussed above does not work, 
in the following we discuss the next simplest possibility of 
coupling two anti-triplets $\bar \phi$ to
the product of two sextet flavons that have the tri-bimaximal alignment,
$\vl \chi^{[p]}_{TB} \vr$ with $p=0,1,2$.
We then prevent the flavon anti-triplets from coupling to
the sextet flavon $\chi^{}_{\mathrm{top}}$ by means of a particular
messenger sector and a $U(1)$
symmetry as discussed in Appendix~\ref{app-control}.
The
relevant scalar potential for the anti-triplets then reads
\bea
V&\!\!=\!\!&-m^2 \!\cdot \!\sum_{i=1}^3{\bar \phi_i}^\dagger {\bar \phi^i}_{} +
\lambda \left(\sum_{i=1}^3 {\bar \phi_i}^\dagger {\bar \phi^i}_{} \right)^2\!+
\sum_{\alpha=1}^3\sum_{i,j=1}^3\sum_{k,l=1}^6
c^{\,\alpha , i}_{\,~~jkl} \, {\bar \phi_i}^\dagger {\bar \phi^j}_{}
 \vl {\chi^{[p]}_{TB}} \vr_k^\dagger \vl \chi^{[p']}_{TB}\vr_l^{} \
 .\label{triplet-align}
\eea
Here the index $\alpha$ labels the three different invariants of the
$PSL_2(7)$ product
$$
\underbrace{(\,{\bf 3} ~\otimes~ {\bf  \ol 3}\,)}_{{\bf 1} \,\oplus\, {\bf 8}}
~\otimes\:
\underbrace{(\,{\bf 6} ~\otimes~
{\bf  6}\,)}_{{\bf 1} \,\oplus\, 2 \cdot {\bf 6} \,\oplus\,{\bf 7} \,\oplus\, 2\cdot {\bf 8}} \ ,
$$
with the index structure of $c^{\,\alpha , i}_{\,~~jkl}$ being defined by the
Clebsch Gordan coefficients of the corresponding Kronecker products.
The contraction to the singlet yields a term of the form
\be
\Delta V_0  ~=~ \alpha_0 \cdot \sum_{i=1}^3 \bar \phi^\dagger_i \bar
\phi^i_{}\ ,
\ee
where $\alpha_0$ includes the VEVs of the sextet flavons. This contribution to
the potential does not constrain the anti-triplet alignment. The situation
changes for the other two invariants obtained form the contractions to the
octet. Inserting the alignments $\vl  \chi^{[p]}_{TB} \vr$ and $\vl
\chi^{[p']}_{TB} \vr$, with $p,p'=0,1,2$, into
Eqs.~(\ref{Delta-sym},\ref{Delta-antisym}) we find that, in general, only the
third and the forth components of the octet of ${\bf 3 \otimes \ol   3}$ in
Eq.~(\ref{33b-octet}) can be projected out in the potential.  However, with
$p=p'$, i.e. identical sextets entering in the potential, both terms vanish
identically. Only in the case where $p\neq p'$ we get non-zero contributions
to the potential.  Choosing for example $p=1$ and $p'=2$, the symmetric octet
of Eq.~(\ref{Delta-sym}) is proportional to $(0,0,0,1,0,0,0,0)$ while the
antisymmetric octet of Eq.~(\ref{Delta-antisym}) is proportional to
$(0,0,1,0,0,0,0,0)$. We thus find two additional independent terms
\bea
\Delta V_s &=& \alpha_s \left[
 \bar \phi^{\dagger}_1 (\bar\phi_{}^2 + \bar \phi_{}^3)
+ \bar\phi^{\dagger}_2(\bar \phi_{}^1 + \bar \phi_{}^3)
+ \bar\phi^{\dagger}_3 (\bar \phi_{}^1 + \bar \phi_{}^2)       \right] \ ,\\
\Delta V_a &=& \alpha_a \left[
\bar\phi^{\dagger}_1 (-2 \bar \phi_{}^1 + \bar
\phi_{}^2 + \bar \phi_{}^3)
+\bar\phi^{\dagger}_2 (\bar \phi_{}^1 + \bar \phi_{}^2 - 2 \bar
\phi_{}^3)
+ \bar\phi^{\dagger}_3 (\bar \phi_{}^1 - 2 \bar \phi_{}^2 +
\bar \phi_{}^3)       \right] \ .~~
\eea
Combining these two terms linearly with each other and with
$\Delta V_0$ we can define two new independent terms which determine the
anti-triplet alignment\footnote{Explicitly, $\Delta V_1 /\alpha_1
= \Delta V_s /\alpha_s +\Delta V_0 /\alpha_0$ and
$ \Delta V_2 /\alpha_2 = \frac{1}{3} (\Delta V_a /\alpha_a - \Delta V_s /\alpha_s +2 \Delta V_0 /\alpha_0)$.}
\bea
\Delta V_1 &=& \alpha_1
( \bar \phi^{\dagger}_1 + \bar\phi^{\dagger}_2 + \bar\phi^{\dagger}_3 )
( \bar \phi_{}^1 + \bar \phi_{}^2 + \bar \phi_{}^3 ) \ ,\label{triplet-align1}\\
\Delta V_2 &=& \alpha_2 ( \bar\phi^{\dagger}_2 - \bar\phi^{\dagger}_3)
(\bar \phi_{}^2-\bar \phi_{}^3)\ .\label{triplet-align2}
\eea
These two terms of the scalar potential are at the core of the discussion of
the anti-triplet alignment. Choosing the values of $\alpha_1$ and $\alpha_2$
appropriately gives rise to a potential which is minimised by an alignment of
the anti-triplets required to generate the first and second family Yukawa
couplings. As an aside we note that the two independent terms of
Eqs.~(\ref{triplet-align1},\ref{triplet-align2}) can be obtained similarly from
Eq.~(\ref{triplet-align}) using tri-bimaximal flavon sextets with $p=0$ and
$p'=1~\mathrm{or}~2$.

Let us first consider the two corrections to the potential {\it individually}. For
positive $\alpha_1$ the minimum of $\Delta V_1$ is zero. This entails a
partial alignment of the form
$$
\alpha_1 > 0: ~~~ \vl \bar \phi \vr
\propto \begin{pmatrix} x \\ y \\ -x-y \end{pmatrix} \ ,
$$
where $x,y \in \mathbb{C}$ remain undetermined. In contrast, for negative
$\alpha_1$ the resulting alignment is completely fixed
$$
\alpha_1 < 0: ~~~ \vl \bar \phi \vr
\propto \begin{pmatrix} 1 \\ 1 \\ 1 \end{pmatrix} \ .
$$
Similarly, the potential term $\Delta V_2$ gives rise to the following
structure of the anti-triplet VEVs
$$
\alpha_2 > 0: ~~~ \vl \bar \phi \vr
\propto \begin{pmatrix} x \\ y \\ y \end{pmatrix} \ , \qquad
\alpha_2 < 0: ~~~ \vl \bar \phi \vr
\propto \begin{pmatrix} 0 \\ 1 \\ -1 \end{pmatrix} \ .
$$
Remarkably, the alignments derived from both terms $\Delta V_1$ and $\Delta
V_2$ individually can be made compatible by choosing the signs of $\alpha_1$
and $\alpha_2$ according to the following combinations
\bea
\alpha_1 > 0\,,~ \alpha_2 < 0: ~~ \vl \bar \phi \vr
\propto \begin{pmatrix} 0 \\ 1 \\ -1 \end{pmatrix} \ ,  \qquad
\alpha_1 < 0\,,~ \alpha_2 > 0: ~~ \vl \bar \phi \vr
\propto \begin{pmatrix} 1 \\ 1 \\ 1 \end{pmatrix} \ .
\eea
Thus it is possible to have an anti-triplet flavon field $\bar\phi_{23}$ whose
VEV becomes aligned along $(0,1,-1)$, while another flavon field
$\bar\phi_{123}$ ends up with the alignment $(1,1,1)$.

\section{Conclusion}
\cleqn
In this paper we have constructed a realistic SUSY GUT of Flavour based on
$PSL_2(7)\times SO(10)$, where the quarks and leptons in the ${\bf 16}$ of
$SO(10)$ are assigned to the complex triplet representation of $PSL_2(7)$,
while the flavons are assigned to a combination of sextets and anti-triplets
of $PSL_2(7)$. It represents the first model based on the finite group
$PSL_2(7)$, which is the smallest simple group that contains both complex
triplet and (real) sextet representations. This group seems particularly well
suited to $SO(10)$ since the sextets may be used to provide the large third
family Yukawa coupling, as well as type II neutrino masses. Furthermore
$PSL_2(7)$ contains $S_4$ as a subgroup, and this allows the possibility of
explaining TB neutrino mixing in a direct way, by preserving the generators
$S,U$ of $PSL_2(7)$ in the neutrino sector, which become identified as the
neutrino flavour symmetry. There are very few models that can account for TB
neutrino mixing using $SO(10)$ and this is the first model which can do this
directly.

Using a $D$-term vacuum alignment mechanism, we have shown how the flavon
sextets of $PSL_2(7)$ can be aligned along the 3-3 direction leading to the
third family Yukawa couplings. Such sextets aligned along the 3-3 direction
are ideally suited for giving a universal contribution to the 3-3 Yukawa
coupling at the lowest possible one-flavon order in $SO(10)$ models, allowing
a sizeable universal top-bottom-tau Yukawa coupling. 
We emphasise that the fact that
a successful potential can be found with particular
values of parameters which can lead to the desired 3-3 vacuum alignment
of the sextet flavons
is highly non-trivial and this is not the case in general for other alignments.
However, in order to realise the
flavon sextet potential that yields an alignment along this 3-3 direction, it
is necessary to assume certain relations amongst the parameters
of the potential. These relations could in principle emerge from a higher symmetry,
beyond $PSL_2(7)$, although this takes us beyond the scope of the present paper,
though it should be the subject of future investigation.

Other sextets are aligned along the neutrino flavour symmetry preserving
directions in an even more natural way, without requiring any relations between
the parameters of the potential, and such alignments suggest
the possibility of TB neutrino mixing via a type
II see-saw mechanism. We have explored the phenomenological consequences of 
such a type II see-saw mechanism 
and obtained statistical predictions for neutrinoless double beta decay
and neutrino masses in cosmology. The distributions of randomly generated
points exhibit a very broad peak at about $\sim 0.05~\mathrm{eV}$ with
significant tails out to about $\sim 0.4~\mathrm{eV}$ in both
$m_{\mathrm{min}}$ and $m_{ee}$.

Anti-triplet flavons are also introduced and aligned against the pre-aligned
TB sextet flavons, in order to give the remaining structure of the charged
fermion mass matrices. In principle these may also be used to account for
neutrino masses and TB mixing  via a type I see-saw mechanism as in
\cite{King:2003rf}, but since this is well known we have focused on the new
type II possibility opened up by having sextets in the model. This also avoids
the use of operators with zero Clebsch factors in order to provide the
necessary suppression in the type I Dirac neutrino sector
\cite{King:2003rf}. Nevertheless, the anti-triplet flavons are instrumental
in giving the successful mass matrices in Eq.~(\ref{dir-matrix2}) via the
assumption of two different messenger mass scales in the up and down
sectors. In this model the mass matrices are achieved in a very natural way
since, with the inclusion of the singlet flavon $\xi$, the first row and
column is cubic in the messenger mass, while the 2-3 block is quadratic and
the 3-3 element involves the universal Higgs messenger mass $M_H$.

In conclusion the SUSY GUT of Flavour based on $PSL_2(7)\times SO(10)$
leads to a very elegant model, combining the mathematical beauty of $PSL_2(7)$
with the attractiveness of $SO(10)$ unification, and solving many of the
problems related to achieving successful fermion masses and TB mixing in the
$SO(10)$ framework. The SUSY flavour problem is also solved here exactly
as in  the $SU(3)$ model discussion in \cite{Antusch:2008jf}, since both
models use identical anti-triplet flavon alignments. Finally we emphasise that 
the type II see-saw mechanism in this model is consistent with  
neutrinoless double beta decay right up to the limit of current experiments.

\section*{Acknowledgments}
We thank Ben Allanach, John Duncan, Michal Malinsky and Alexander Merle for
helpful discussions. SFK and CL acknowledge support from STFC Rolling Grant
ST/G000557/1. SFK acknowledges the support of a Royal Society - Leverhulme
Trust Senior Research Fellowship.

\begin{appendix}
~\\[5mm]
\noindent{\bf{\Large{Appendix}}}
\section{\label{app-induced}The induced VEV}
\cleqn
In $SO(10)$ the induced VEV of the $\Delta_{\ol{126}}$ field in
Fig.~\ref{maj-fig}\,(a) is obtained by replacing the $\Delta_{\ol{126}}$ leg
of the diagram by
\begin{center}
\includegraphics[width=3cm]{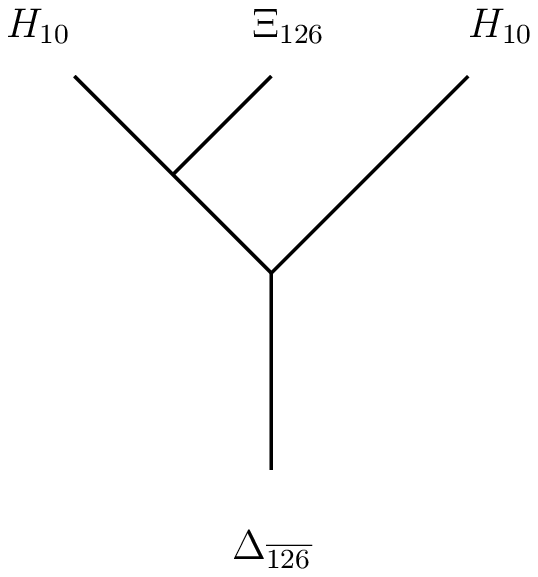}
\end{center}
The relevant components of the $SO(10)$ representations are
\be
H_{10} ~\rightarrow~ ({\bf 15} , {\bf 2} , {\bf 2}) \ , \qquad
\Xi_{126} ~\rightarrow~ ({\bf \ol{10}},{\bf 1},{\bf 3})  \ ,
\ee
with $H_{10}$ acquiring a VEV at the electroweak scale while $\Xi_{126}$ gets
a GUT scale VEV. Assuming the messengers in the complete diagram to have a
mass of order $M$, the induced VEV can be calculated to be
\be
\frac{{\vl \Xi_{126}\vr_{({\bf \ol{10}},{\bf 1},{\bf 3})}}  \:
{\vl H_{10} \vr_{({\bf {1}},{\bf 2},{\bf 2})}} \:
{\vl H_{10} \vr_{({\bf {1}},{\bf 2},{\bf 2})}}}{M^2}
~\sim~ \frac{v^2_u}{M} \ ,\label{ind-eq}
\ee
where, for simplicity, we have taken
$\vl \Xi_{126}\vr_{({\bf \ol{10}},{\bf 1},{\bf 3})} \sim M$.
It is clear from Eq.~(\ref{ind-eq}) that the resulting contribution to the
neutrino masses corresponds to the type II see-saw mechanism because  the two
left-handed doublets of the $H_{10}$ pair combine to an
$SU(2)_L$ triplet, whereas the two right-handed doublets are contracted with
the right-handed triplet of $\Xi_{126}$ to a singlet under $SU(2)_R$. The
relevant component of the $\Delta_{\ol{126}}$ messenger is thus an
$SU(2)_L$ triplet.

\section{\label{app-control}Controlling the flavon potential}
\cleqn
In Section~\ref{vac-sec} we discussed the terms of the flavon potential that
are required to obtain the alignments that generate the Majorana and Yukawa
couplings. There we encountered the following four types of couplings
\be
\chi^\dagger_{\mathrm{top}} \chi^{}_{\mathrm{top}} \chi^\dagger_{\mathrm{top}}
\chi^{}_{\mathrm{top}}  \, , \quad
 \chi^{[p] \, \dagger}_{TB}  \chi^{[p]}_{TB} \chi^{[p] \,
  \dagger}_{TB}  \chi^{[p]}_{TB} \, , \quad
\bar\phi_{23}^\dagger \bar\phi_{23}^{}  \chi^{[p] \, \dagger}_{TB}
\chi^{[p']}_{TB} \, , \quad
\bar\phi_{123}^\dagger \bar\phi_{123}^{}  \chi^{[p] \, \dagger}_{TB}
\chi^{[p']}_{TB} \, ,\label{flavo-po}
\ee
which involve four sextet and two anti-triplet flavon fields. As we have already
discussed the contractions that yield $PSL_2(7)$ invariants, we suppress all
indices in Eq.~(\ref{flavo-po}). At this point it is important to observe
that the suggested vacuum alignment is based on the absence of terms like
\be
\chi^\dagger_{\mathrm{top}} \chi^{}_{\mathrm{top}}   \chi^{[p] \, \dagger}_{TB}  \chi^{[p]}_{TB}  \ , ~\quad
 \chi^{[p] \, \dagger}_{TB}  \chi^{[p]}_{TB}  \chi^{[p'] \,
  \dagger}_{TB}  \chi^{[p']}_{TB} \ , ~\quad
\bar\phi_{23}^\dagger \bar\phi_{23}^{} \chi^{\dagger}_{\mathrm{top}}
\chi^{}_{\mathrm{top}} \ , ~\quad
\bar\phi_{123}^\dagger \bar\phi_{123}^{} \chi^{\dagger}_{\mathrm{top}}
\chi^{}_{\mathrm{top}} \ ,\label{flavo-mix}
\ee
that -- at the effective level -- cannot be forbidden by symmetries alone.
It is therefore necessary to resort to a particular messenger sector. A
neutral messenger would automatically give rise to diagrams such as
\begin{center}
\includegraphics[height=35mm]{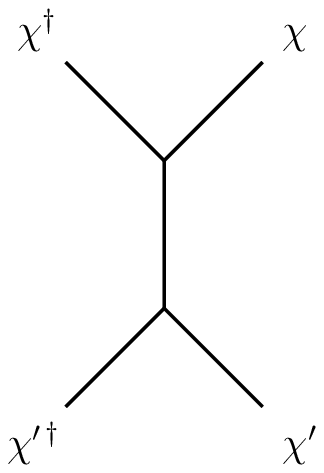} \hspace{3cm}
\includegraphics[height=35mm]{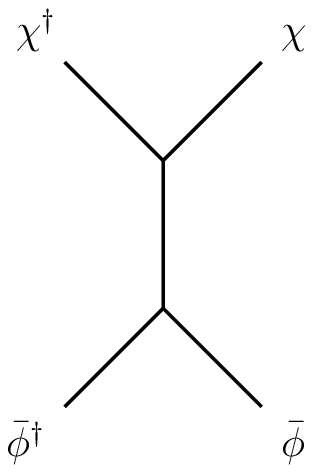}
\end{center}
which cannot distinguish between the structure of the terms in
Eq.~(\ref{flavo-po}) and  Eq.~(\ref{flavo-mix}).  Therefore the messengers
must be charged under additional symmetries. Accordingly, a possible way to
forbid the terms of Eq.~(\ref{flavo-mix}) is given by diagrams of type
\begin{center}
\includegraphics[width=35mm]{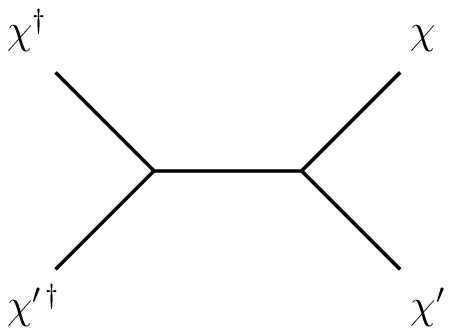} \hspace{1.9cm}
\includegraphics[width=35mm]{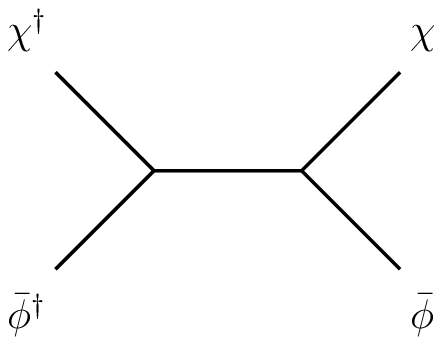}
\end{center}
where the charge of the respective messengers determines whether or not $\chi'
= \chi$ is allowed as well as which sextet flavons can couple to the
anti-triplet flavons. Most notably we need to separate the top sextet
from the sextets of tri-bimaximal type. This is achieved easily by assigning
different $U(1)$ charges $q$ to the flavons $\chi^{}_{\mathrm{top}} $ and
$ \chi^{[p]}_{TB}$. Choosing for instance
\be
q(\chi^{}_{\mathrm{top}}) = -1 \ , \qquad
q(\chi^{[p]}_{TB}) = -2 \ , \qquad
q(\bar\phi^{}_{23}) = -2 \ , \qquad
q(\bar\phi^{}_{123}) = 4 \ ,
\ee
one can generate the operators of Eq.~(\ref{flavo-po}) using messengers with
charges $q=2,4,4,-2$, respectively, while the first, third and fourth
term of Eq.~(\ref{flavo-mix}) would require messengers with odd $U(1)$
charge. In the absence of such messengers the top sextet flavon cannot mix with
the tri-bimaximal one.

Furthermore, we also need to forbid the second term of
Eq.~(\ref{flavo-mix}), i.e. the mixing among the three tri-bimaximal
flavon sextets. For this purpose we introduce a separate symmetry, $U(1)'$,
which distinguishes between $ \chi^{[p]}_{TB}$, with $p=0,1,2$.
One possible set of $U(1)'$ charges could be
\be
q'(\chi^{[0]}_{TB}) = 2 \ , \qquad
q'(\chi^{[1]}_{TB}) = 3 \ , \qquad
q'(\chi^{[2]}_{TB}) = 5 \ . \label{q'-values}
\ee
The second term of Eq.~(\ref{flavo-po}), corresponding to three distinct
quartic operators, would arise from messengers with charges $q'=-4,-6,-10$,
respectively. At the same time, the analogous mixing term of
Eq.~(\ref{flavo-mix}) would require messengers with either $q'=-5,-7,-8$ or
$q'=1,2,3$.

In the construction of the complete model it is necessary to introduce three
$PSL_2(7)$ singlet flavons $\zeta^{[p]}_{}$. These are associated with the three
tri-bimaximal sextet flavons $ \chi^{[p]}_{TB}$ and carry opposite $U(1)'$
charge. Then the $\chi^{[p]}_{TB}$ legs of the diagrams in Fig.~\ref{maj-fig}
need to be replaced by
\begin{center}
\includegraphics[width=33mm]{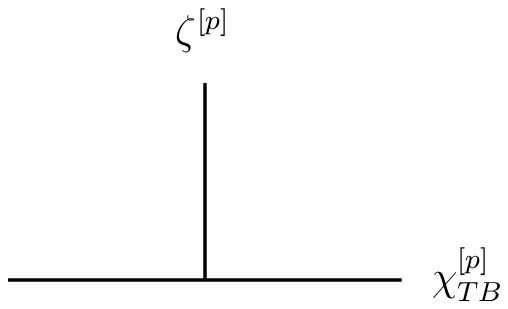}
\end{center}
With this completion, the only particles that are charged under the $U(1)'$
symmetry are $\chi^{[p]}_{TB}$, $\zeta^{[p]}_{}$ and the above mentioned
messengers with $q'=-4,-6,-10$.

\section{\label{octet}The octet representation}
\cleqn

Just like in $SU(3)$, the octet can be constructed from the product ${\bf 3
\otimes \ol 3}$, see for instance Ref.~\cite{PSLgroup}. The resulting
$8\times 8$ matrices for the $PSL_2(7)$ generators are real, but none is
diagonal. Since we are interested in the combination $({\bf 6\otimes 6})
\rightarrow {\bf 8}$, it is convenient to perform a similarity transformation
to a basis in which --~analogous to the sextet (see \cite{King:2009mk})~--
the generators $\m S^{[{\bf 8}]}$ and $\m U^{[{\bf 8}]}$ are diagonal. We obtain
\bea
\m S^{[{\bf 8}]} &=&
\text{Diag}\,(-1\,,\,-1\,,\,1\,,\,1\,,\,-1\,,\,-1\,,\,1\,,\,1) \
, \label{oct-S}\\[4mm]
\m T^{[{\bf 8}]} &=&
\frac{1}{2} \begin{pmatrix}
1&0&\sqrt{2}&0&0&1&0&0 \\
0&-1&0&0&1&0&\sqrt{2}&0 \\
\sqrt{2}&0&0&0&0&-\sqrt{2}&0&0 \\
0&0&0&-1&0&0&0&-\sqrt{3} \\
0&-1&0&0&1&0&-\sqrt{2}&0 \\
-1&0&\sqrt{2}&0&0&-1&0&0 \\
0&-\sqrt{2}&0&0&-\sqrt{2}&0&0&0 \\
0&0&0&\sqrt{3}&0&0&0&-1
\end{pmatrix} \ ,\label{oct-T}\\[4mm]
\m U^{[{\bf 8}]} &=&
\text{Diag}\,(1\,,\,1\,,\,1\,,\,1\,,\,-1\,,\,-1\,,\,-1\,,\,-1) \
,\label{oct-U} \\[4mm]
\m V^{[{\bf 8}]} &=&
\frac{1}{4} \begin{pmatrix}
-3&\sqrt{7}&0&0&0&0&0&0 \\
\sqrt{7}&3&0&0&0&0&0&0 \\
0&0&2&2\sqrt{3}&0&0&0&0 \\
0&0&2\sqrt{3}&-2&0&0&0&0 \\
0&0&0&0&0&0&-\sqrt{2}&-\sqrt{14} \\
0&0&0&0&0&0&-\sqrt{14}&\sqrt{2} \\
0&0&0&0&-\sqrt{2}&-\sqrt{14}&0&0 \\
0&0&0&0&-\sqrt{14}&\sqrt{2}&0&0
\end{pmatrix} \label{oct-V} \ .
\eea
With the identification
$$
A ~=~
\left[\m T^{[{\bf 8}]}\m U^{[{\bf 8}]}\m S^{[{\bf 8}]}\m (T^{[{\bf
 8}]})^2\, \m U^{[{\bf 8}]}\right]^{-1}
\, \m V^{[{\bf 8}]} \,
\left[\m T^{[{\bf 8}]}\m U^{[{\bf 8}]}\m S^{[{\bf 8}]}\m (T^{[{\bf 8}]})^2\, \m
U^{[{\bf 8}]}\right] \ , ~\quad
B ~=~ \m T^{[{\bf 8}]} \ ,
$$
one can easily check that the presentation of $PSL_2(7)$ is satisfied.
The symmetric octet~$\Omega$ of Eq.~(\ref{Delta}) is given in the basis
of Eqs.~(\ref{oct-S}-\ref{oct-V}). More generally there are two independent
octets derived from the product of two sextets which we take in the basis of
\cite{King:2009mk}. They are
\be
\begin{pmatrix}
\sqrt{3} \chi^{}_2\chi'_3 + \chi^{}_1\chi'_4 -2\sqrt{6} \chi^{}_1\chi'_5 +2\sqrt{14}
\chi^{}_1\chi'_6 \\[1mm]
 \sqrt{21}\chi^{}_2\chi'_3 -3\sqrt{7} \chi^{}_1\chi'_4 \\[1mm]
-\sqrt{6}\chi^{}_1\chi'_1 +\sqrt{6}\chi^{}_2\chi'_2 -2\chi^{}_4\chi'_5
+2\sqrt{14}\chi^{}_5\chi'_6 \\[1mm]
\!\!- \sqrt{2}\chi^{}_1\chi'_1  -  \sqrt{2}\chi^{}_2\chi'_2
+2\sqrt{2}\chi^{}_3\chi'_3-2\sqrt{2}\chi^{}_4\chi'_4+2\sqrt{2}\chi^{}_5\chi'_5-2
\sqrt{7}\chi^{}_4\chi'_6 \!\! \\[1mm]
-\sqrt{21}\chi^{}_1\chi'_3  +3\sqrt{7} \chi^{}_2\chi'_4\\[1mm]
\sqrt{3}
\chi^{}_1\chi'_3+\chi^{}_2\chi'_4+2\sqrt{6}\chi^{}_2\chi'_5+2\sqrt{14}\chi^{}_2\chi'_6\\[1mm]
-2\sqrt{21}\chi^{}_3\chi'_5\\[1mm]
2\sqrt{6}\chi^{}_1\chi'_2-4\sqrt{2}\chi^{}_3\chi'_4+2\sqrt{7}\chi^{}_3\chi'_6
\end{pmatrix} \!+ (\chi \leftrightarrow \chi'),~~\label{Delta-sym}
\ee
\be
\begin{pmatrix}
-\sqrt{21} \chi^{}_2\chi'_3 - \sqrt{7} \chi^{}_1\chi'_4 -2\sqrt{2}
\chi^{}_1\chi'_6 \\[1mm]
 \sqrt{3}\chi^{}_2\chi'_3 -3 \chi^{}_1\chi'_4 -2\sqrt{6} \chi^{}_1\chi'_5\\[1mm]
 -2\sqrt{7}\chi^{}_4\chi'_5 -2\sqrt{2}\chi^{}_5\chi'_6 \\[1mm]
-6\chi^{}_4\chi'_6  \\[1mm]
-\sqrt{3}\chi^{}_1\chi'_3  +3 \chi^{}_2\chi'_4 -2\sqrt{6} \chi^{}_2 \chi'_5\\[1mm]
-\sqrt{21} \chi^{}_1\chi'_3-\sqrt{7} \chi^{}_2\chi'_4 -2\sqrt{2}\chi^{}_2\chi'_6\\[1mm]
2\sqrt{6} \chi^{}_1\chi'_2+ 2\sqrt{3}\chi^{}_3\chi'_5\\[1mm]
6 \chi^{}_3\chi'_6
\end{pmatrix} \!- (\chi \leftrightarrow \chi').~~\label{Delta-antisym}
\ee
Obviously, the antisymmetric one only exists for $\chi \neq \chi'$. Adopting
the triplet basis of \cite{King:2009mk}, the octet of the product ${\bf
  3\otimes \ol 3}$ reads
\be
\begin{pmatrix}
\phi^{}_1 (4 \bar \phi_{}^1 + \bar \phi_{}^2 + \bar
\phi_{}^3) + (\phi^{}_2 + \phi^{}_3) (\bar \phi_{}^1 - 2 \bar
\phi_{}^2 -2  \bar \phi_{}^3)\\
- 3 i [ \phi^{}_1 (\bar \phi_{}^2 + \bar \phi_{}^3 )
- (\phi^{}_2 + \phi^{}_3 ) \bar \phi_{}^1   ]\\
-\sqrt{2} [
\phi^{}_1 (-2 \bar \phi_{}^1 + \bar
\phi_{}^2 + \bar \phi_{}^3)
+\phi^{}_2 (\bar \phi_{}^1 + \bar \phi_{}^2 - 2 \bar
\phi_{}^3)
+ \phi^{}_3 (\bar \phi_{}^1 - 2 \bar \phi_{}^2 +
\bar \phi_{}^3) ] \\
\sqrt{6} [
 \phi^{}_1 (\bar\phi_{}^2 + \bar \phi_{}^3)
+ \phi^{}_2(\bar \phi_{}^1 + \bar \phi_{}^3)
+ \phi^{}_3 (\bar \phi_{}^1 + \bar \phi_{}^2) ] \\
\sqrt{3}  [
\phi^{}_1 (\bar \phi_{}^2  -  \bar \phi_{}^3 )
+ \phi^{}_2 (\bar \phi_{}^1 + 2 \bar \phi_{}^2)
-\phi^{}_3 (\bar \phi_{}^1  + 2  \bar \phi_{}^3  ) ]\\
i\sqrt{3} [
\phi^{}_1 (\bar \phi_{}^2 - \bar \phi_{}^3)
- \phi^{}_2 (\bar\phi_{}^1 + 2 \bar \phi_{}^3)
+\phi^{}_3 (\bar \phi_{}^1 + 2 \bar \phi_{}^2) ] \\
\sqrt{6} [
\phi^{}_1(\bar \phi_{}^2 - \bar \phi_{}^3)
+\phi^{}_2 (\bar \phi_{}^1 - \bar \phi_{}^2)
- \phi^{}_3 (\bar\phi_{}^1 - \bar \phi_{}^3)  ] \\
-i\sqrt{6}  [
\phi^{}_1 (\bar\phi_{}^2 - \bar \phi_{}^3)
-\phi^{}_2 (\bar \phi_{}^1 - \bar \phi_{}^3)
+\phi^{}_3 (\bar \phi_{}^1 - \bar \phi_{}^2) ]
\end{pmatrix}. \label{33b-octet}
\ee

\end{appendix}



\end{document}